\documentclass[%
 reprint,
superscriptaddress,
nofootinbib,
 amsmath,amssymb,
 aps,
]{revtex4-2}
\usepackage{graphicx}
\usepackage{dcolumn}
\usepackage{bm}
\usepackage{hyperref}
\hypersetup{
    colorlinks,
    citecolor=blue,
    linkcolor=blue,
    urlcolor=blue,
}

\begin{document}

\title{Constraints on Lorentz invariance violation from the LHAASO observation of GRB 221009A }
\author{Yu-Ming Yang}
\email{yangyuming@ihep.ac.cn}
 \affiliation{%
 Key Laboratory of Particle Astrophysics, Institute of High Energy Physics, Chinese Academy of Sciences, Beijing 100049, China}
\affiliation{
 University of Chinese Academy of Sciences, Beijing 100049, China 
}%
\author{Xiao-Jun Bi}
\email{bixj@ihep.ac.cn}
\affiliation{%
 Key Laboratory of Particle Astrophysics, Institute of High Energy Physics, Chinese Academy of Sciences, Beijing 100049, China}
\affiliation{
 University of Chinese Academy of Sciences, Beijing 100049, China 
}%
\author{Peng-Fei Yin}
\email{yinpf@ihep.ac.cn}
\affiliation{%
 Key Laboratory of Particle Astrophysics, Institute of High Energy Physics, Chinese Academy of Sciences, Beijing 100049, China}

\begin{abstract}
In some quantum gravity (QG) theories, Lorentz symmetry may be broken above the Planck scale. The Lorentz invariance violation (LIV) may induce observable effects at low energies and be detected at high energy astrophysical measurements.
The Large High Altitude Air Shower Observatory (LHAASO) has detected the onset, rise, and decay phases of the afterglow of GRB 221009A, covering a wide energy range of photons approximately from 0.2 to 13 TeV. This observation provides an excellent opportunity to study the Lorentz invariance violation effect. In this study, we simultaneously utilize the data from the KM2A and WCDA detectors of LHAASO, and apply two event by event methods, namely the pair view method and maximum likelihood method, to investigate LIV. We obtain stringent constraints on the QG energy scale. For instance, through the maximum likelihood method, we determine the $95\%$ confidence level lower limits to be $E_{\text{QG},1} > 14.7\, (6.5)\times 10^{19}$GeV  for the subluminal (superluminal) scenario of $n=1$, and $E_{\text{QG},2} > 12.0 \,(7.2) \times 10^{11}$GeV for the subluminal (superluminal) scenario of $n=2$. We find that the rapid rise and slow decay behaviors of the afterglow can impose strong constraints on the subluminal scenario, while the constraints are weaker for the superluminal scenario.

\end{abstract}


\maketitle

\section{Introduction\label{introduction}}

Quantum field theory and general relativity are the two cornerstones of modern physics. Quantum field theory has achieved remarkable success in elucidating electromagnetic, strong, and weak interactions, serving as the foundation of the standard model of particle physics. In parallel, general relativity has made great achievements in elucidating gravity, forecasting the existence of black holes, and establishing the $\Lambda$CDM cosmology. Nevertheless, it is widely acknowledged that discrepancies exist between these two theories \cite{Burgess_2004,PhysRevD.10.411,Addazi_2022}. In extreme environments, such as the core of black holes or in the vicinity of the singularity of the Big Bang, the quantum effects of gravity become non-negligible \cite{Bonanno_1999,Bosma_2019,Gambini_2014}. Consequently, there arises a need for a unified theory of quantum gravity (QG) that harmoniously integrates quantum field theory and general relativity, driven by both the perspective of theoretical self-consistency and the necessity to elucidate certain phenomena.

The breaking of Lorentz symmetry serves as a characteristic feature of numerous QG theories \cite{Amelino_Camelia_2013,PhysRevD.39.683,Carroll_2001,Addazi_2022}. One method for formulating QG theories is the top-down approach, which involves the construction of theories purely from a theoretical standpoint, such as string theory \cite{becker2006string,Mukhi_2011,Mattingly_2005}, asymptotically-safe gravity \cite{PhysRevD.57.971,percacci2017introduction,bonanno2020critical}, and others. Many of these theories predict the breaking of Lorentz symmetry. Conversely, the bottom-up approach considers the effects of Lorentz symmetry breaking, for example, through the standard model extension \cite{Colladay_1997,Kosteleck__2009} or by altering the form of Lorentz transformations like doubly special relativity \cite{AMELINO_CAMELIA_2002,Kowalski_Glikman}. This approach utilizes observations to impose constraints on these effects, thereby guiding the development of QG theories.

The Lorentz invariance violation (LIV) can alter the dispersion relation of photons in vacuum, and lead to numerous observable effects such as energy-dependent speed of light \cite{1999PhRvL..83.2108B,Albert_2008,Aharonian_2008,Abramowski_2011,Ellis_2003,chrétien2015constraining,2017,piran2023lorentz,Pasumarti_2023}, photon decay or split \cite{LHAASO:2021opi,Astapov_2019,Albert_2020},  pair-production threshold anomaly \cite{Lang_2019,Abdalla_2019,Li_2023}, and others. Notable advancements have been achieved in these areas in in the past years \cite{heros2022cosmic}. However, the application of some of these phenomena is subject to certain limitations. For instance, the photon decay effect is only permissible in the superluminal scenario, thereby restricting related studies to setting constraints solely on this scenario. Conversely, the energy dependence of the speed of light offers a direct approach to investigating LIV, which enables the simultaneous exploration of both superluminal and subluminal scenarios.

Research on the energy dependence of the speed of light involves studying the time-of-flight of photons, which imposes strict requirements on the light source. Previous studies have utilized various sources such as pulsars \cite{2017,chrétien2015constraining}, active galactic nuclei
\cite{1999PhRvL..83.2108B,Albert_2008,Abramowski_2011},  gamma ray bursts (GRBs) \cite{Amelino_Camelia_1998,Vasileiou_2013,Acciari_2020}. Additionally, a wide variety of analysis methods have been developed, including binned methods such as the modified cross correlation function \cite{Li_2004,Aharonian_2008}and continuous wavelet transform \cite{Ellis_2003}, and unbinned methods such as pair view (PV) \cite{Vasileiou_2013} and maximum likelihood (ML) \cite{Mart_nez_2009} methods. 

In this study, we utilize the observation of the bright gamma-ray burst, GRB 221009A, from the Large High Altitude Air Shower Observatory  (LHAASO) \cite{2023,cao2023high} to set constraints on the QG energy scale. In order to fully exploit the information carried by the observation, we employ two distinct event by event methods, namely PV and ML methods.  In the PV analysis, we simultaneously utilize data from two detectors of LHAASO, namely the Water Cherenkov Detector Array (WCDA) and the Kilometer Squared Array (KM2A), to increase the statistical significance and broaden the energy range of photons. 
Meanwhile, in the ML analysis, we employ the light curve template obtained by fitting to the photons from the lowest $N_\text{hit}$ bin of WCDA, and use high-energy photons from KM2A to calculate the likelihood function. Our study encompasses three different intrinsic energy spectral shapes and the corresponding extragalactic background light (EBL) absorption models, and we discuss their impact on the constraint results. 

The paper is organized as follows. In Sec.~\ref{TC} we  introduce the theoretical basis of LIV and related conventions, while in Sec.~\ref{GRB} we briefly present the observations of GRB 221009A by LHAASO used in this study. The analyses with the PV and the ML methods are performed in Sec.~\ref{PVM} and Sec.~\ref{MLM}, respectively. Finally, we provide a summary of the study in Sec.~\ref{Sum}.

\section{Theoretical background and Conventions\label{TC}}
The presence of LIV modifies the dispersion relationship of photons in vacuum, which can be phenomenologically expressed as \cite{Acciari_2020} 
\begin{equation}
    E^2=p^2c^2\left[1-\sum_{n=1}^\infty \mathcal{S}\left(\frac{E}{E_{\text{QG},n}}\right)^n\right],
\end{equation}
where $\mathcal{S}=+1$ and $-1$ refer to  the subluminal and superluminal scenarios, respectively, and $E_{\text{QG},n}$ represents the QG energy scale. At this scale, the Compton wavelength and Schwarzschild radius of a particle with a mass $m$ are of the same order of magnitude, and the effects of QG cannot be ignored. With $\hbar/mc\simeq Gm/c^2$, the QG energy scale can be expected to be at the order of the Planck mass $M_\text{PL} \equiv \sqrt{\hbar c/G}\simeq  1.22\times 10^{19}\text{GeV}$ \footnote{It is worth mentioning that the QG energy scale permits for a departure from the Planck mass scales \cite{Addazi_2022,Amelino_Camelia_2013}. For instance, in string theory, the higher-dimensional gravitational energy
scale, i.e. the scale at which the QG effects cannot be ignored, could be much lower than the Planck mass.}, which is significantly higher than that achievable in collider experiments or involved in currently known astrophysical processes. Consequently, only the lowest order terms dominate the above expression. In this analysis, we only consider the case where the lowest order term is $n = 1$ or $n = 2$.  The group velocity of light in vacuum can be deduced from the above dispersion relationship as 
\begin{equation}
    v_{\gamma}(E)=\frac{\partial E}{\partial p}\simeq c\left[1-\mathcal{S}\frac{n+1}{2}\left(\frac{E}{E_{\text{QG},n}}\right)^n\right].
\end{equation}
This implies that photons of different energies travel at different speeds. When a source at a specific redshift $z_s$ emits two photons with different energies $E_1$ and $E_2$ simultaneously,  there is an energy-dependent difference between the arrival times of the two photons at the Earth 
\begin{equation}
\begin{aligned}
    \Delta t_{LIV}(n,E_2,E_1)&\simeq \mathcal{S}\frac{n+1}{2}\frac{E^n_2-E^n_1}{E^n_{\text{QG},n}}\int_0^{z_s}dz\frac{(1+z)^n}{H(z)}\\
    &\equiv \tau_n(E^n_2-E^n_1),
\end{aligned}
    \label{deltat}
\end{equation}
where $H(z)=H_0\sqrt{\Omega_m(1+z)^3+\Omega_\Lambda}$ is the Hubble parameter at redshift $z$, and $\tau_n$ is a parameter defined for convenience in the PV analysis. The parameters $H_0= 67.36 \; \text{km}\,\text{s}^{-1}\text{Mpc}^{-1}$, $\Omega_m= 0.315$, and $\Omega_\Lambda\simeq 1-\Omega_m =0.685$ are the Hubble parameter, matter density, and dark energy density of the current universe, respectively, with values based on the Planck 2018 results \cite{Planck_2018}. For numerical convenience, we also define two additional parameters as \cite{Acciari_2020} $\eta_1\equiv \mathcal{S} M_\text{PL}/E_{\text{QG},1}$ and $\eta_2\equiv 10^{-16}\mathcal{S}M^2_{\text{PL}}/E^2_{\text{QG},2}$.

The difference between the arrival times of  two photons increases with their energy difference, highlighting the importance of employing sources that encompass a wide energy spectrum in the study of LIV. Nonetheless, as mentioned earlier, due to the currently known astrophysical energies being significantly lower than the Planck energy scale, the aforementioned effect is exceedingly weak. Consequently, to enhance the detectability of this  subtle phenomenon, it is necessary to accumulate the effect by extending the propagation distance of the photons, thereby emphasizing the importance of sources with high redshifts.

It is important to consider that observed photons are not emitted instantaneously; even the photons with the same energies emitted from the source possess an intrinsic time distribution. Generally, the time lag resulting
from LIV would not be significantly greater than that extracted from the data and the source's variability time $t_\text{var}$, at the very least. Consequently, sources exhibiting rapid variability are deemed more suitable for LIV investigations. From Eq.~\ref{deltat}, we can see that the constraints on the QG scale from a specific source are roughly proportional to $t^{-1}_\text{var}$ and $t^{-1/2}_\text{var}$, for the cases with $n=1$ and $n=2$, respectively \cite{Terzi__2021}. 

Celestial objects commonly employed for LIV studies include pulsars, active galactic nuclei, and GRBs, each offering distinct advantages in the aforementioned aspects. GRBs are short-lived, intense bursts of gamma-ray emission into space, resulting from the collapse of massive stars (long GRBs) or the merger of binary compact stars (short GRBs). The specific source utilized in this study is GRB 221009A.

\section{GRB 221009A\label{GRB}}
On October 9th, 2022, GRB 221009A was initially detected by the Fermi-GBM at 13:16:59.99 Universal Time 
\cite{2022GCN.32636....1V,2023ApJ...952L..42L}. This time $T_0$ serves as the reference point in this study, with all subsequent time measurements expressed in seconds. The burst's isotropic-equivalent energy release $E_{\gamma,\text{iso}}$ is estimated to be approximately $10^{55}$ erg, nearly saturating all gamma-ray detectors. The emission of GRB occurs in two stages: the prompt emission and afterglow, with a partial overlap in time. The prompt emission phase exhibits irregular variability and rapid light changes, typically occurring on the order of milliseconds. In contrast, the afterglow phase is characterized by smoother emission and prolonged duration. It is difficult to observe the onset of the afterglow using pointed instruments. However, with a large field of view and near 100$\%$ duty cycle, LHAASO successfully observed this afterglow approximately 6000 s after the initial detection, detecting the onset and rising phases of afterglow for the first time \cite{2023}.
 
Both detectors of LHAASO, namely WCDA and KM2A, observed GRB 221009A. During the first 3000 seconds after $T_0$, the WCDA detector observed approximately 64,000 photons with energies ranging from 0.2 to 7 TeV \cite{2023}.  The resulting light curve is composed of four segments: rapid rise, slow rise, slow decay, and steep decay, with the onset time estimated to be $T^\star =226$ s. Additionally, the WCDA photon spectra of 15 time intervals reveal a power-law spectral index evolving with time, represented by $\gamma(t)=a\log(t-226)+b$ with $a=-0.135\pm 0.028$ and $b=2.579\pm 0.055$. Furthermore, the other detector KM2A observed 142 photons with energies above 3 TeV in the 230-900 s interval \cite{cao2023high}. Due to the limitation of KM2A's energy resolution, the reconstruction of energy depends on the assumption of the input spectral form. In this work, we use the reconstructed energies based on the observed spectral form of the power-law with an exponential cutoff. 

For the large array LHAASO, the photon energy is characterized by the parameter $N_\text{hit}$, which denotes the number of fired cells with a specific normalized charge threshold in an air shower event \cite{2023}.
Due to the low energy resolution of WCDA, the accurate reconstructed energies of WCDA photons are not available here.  The data for the arrival times of WCDA \footnote{\url{https://www.nhepsdc.cn/resource/astro/lhaaso/paper.Science2023.adg9328/}} photons in various $N_\text{hit}$ bins, along with the arrival times and estimated energies of KM2A \footnote{\url{https://opendata.ihep.ac.cn/nhepsdc@ihep.ac.cn\#/Search/SearchDetail?id=455b4cc08c0d4ffea093046e43fcda1b}}  photons, have been available on public data web pages.

The redshift of GRB 221009A is determined through optical observations as $z_s=0.151$ \cite{2022GCN.32648....1D,2022GCN.32686....1C}. Subsequently, the time lag caused by the LIV can be calculated using Eq.~\ref{deltat} as
\begin{equation}
    \begin{aligned}
    &\Delta t_\text{LIV}(\eta_1,E_2,E_1)=5.87\eta_1\left(\frac{E_2}{\text{TeV}}-\frac{E_1}{\text{TeV}}\right)\, \text{s}, \\
    &\Delta t_\text{LIV}(\eta_2,E_2,E_1)=7.77\eta_2\left[\left(\frac{E_2}{\text{TeV}}\right)^2-\left(\frac{E_1}{\text{TeV}}\right)^2\right]\, \text{s} .
    \end{aligned} 
    \label{DeltatLIV}  
\end{equation}

\section{Pair View Analysis\label{PVM}}
In order to fully exploit the information carried by
the KM2A photons, we use two event by event methods in the analysis. The first one is the PV method, which was initially proposed by Vasileiou et al. \cite{Vasileiou_2013} and was used to analyze 4 GRBs detected by Fermi-LAT. Assuming the detector has detected $N$ photons, each with corresponding arrival time $t$ and estimated energy $E_\text{est}$. In the case where the estimated energies of all photons are different from each other, for any pair of the $i$-th and $j$-th photons, we can calculate the spectral lag between them as
\begin{equation}
    l^{(n)}_{i,j}\equiv \frac{t^{(i)}-t^{(j)}}{\left(E^{(i)}_\text{est}\right)^n-\left(E^{(j)}_\text{est}\right)^n}
    \label{lij}.
\end{equation}
The total number of spectral lags is $N(N-1)/2$, and the following discussion is applicable to both the $n=1$ and $n=2$ cases. If all the photons are emitted at the same time, the time lag $t^{(i)}-t^{(j)}$ is entirely induced by the LIV. Using Eq.~\ref{DeltatLIV}, we  directly derive 
\begin{equation}
    \begin{aligned}
        &l^{(1)}_{i,j}=5.87\eta_1\, \text{s}/\text{TeV}, \\
        &l^{(2)}_{i,j}=7.77\eta_2\,\text{s}/\text{TeV}^2.
    \end{aligned}
    \label{lij_eta}
\end{equation}

If we perform histogram statistics on the $N(N-1)/2$ spectral lags for the photons emitted at the same time, they should concentrate on the values given by Eq.~\ref{lij_eta}, which are equal to $\tau_n$. However, in reality, the photons emitted from the source have an intrinsic time distribution. This intrinsic distribution  induces a non-zero width to the distribution of the spectral lags. Consequently, the distribution of these spectral lags should exhibit a peak on a diffuse background, with the position of the peak corresponding to the LIV parameter $\tau_n$.

Since the photons from the slow decay part of the afterglow contribute mainly to the background rather than the peak, we select a subset of photons from LHAASO with arrival times between 230-300 s, which correspond to the phase of fastest variability in the afterglow, in order to enhance the visibility of the peak. This subset includes 64 photons with energies above 3 TeV detected by KM2A, and 3,818 photons in the $N_\text{hit}\in [30, 33)$ bin  detected by WCDA. 

The accurate reconstruction of energies of the photons observed by WCDA is challenging due to its limited energy resolution. However, as indicated by Eq.~\ref{lij}, when $E^{(i)}_\text{est} \gg E^{(j)}_\text{est}$, the time lags would not be significantly affected by the uncertainty of $E^{(j)}_\text{est}$. This is applicable for determining the spectral lags between the high energy KM2A photons and low energy WCDA photons.  
As a simple assumption, all the energies of the 3,818 WCDA photons are assumed to be the median energy $E_0 = 0.35$ TeV of this $N_\text{hit}$ bin. The impact of this simplification on the constraints will be discussed at the end of this section. With the time bin of 0.1 s, many photons may arrive within the same time bin. In order to account for this, we uniformly and randomly distribute each photon's arrival time within its respective 0.1 s time bin.  Then, the spectral lags between any pair of the 64 photons from KM2A data, and the lags between the KM2A photons and WCDA photons are calculated, resulting in a total of $64\times (64-1)/2+64\times 3818 = 246368 $ spectral lags. We conduct histogram statistics on these spectral lags and use the kernel density estimation method to determine the position of the peak $\hat{\tau}_n$ in unbinned data.
\begin{figure}[htbp]
    \includegraphics[width=\linewidth]{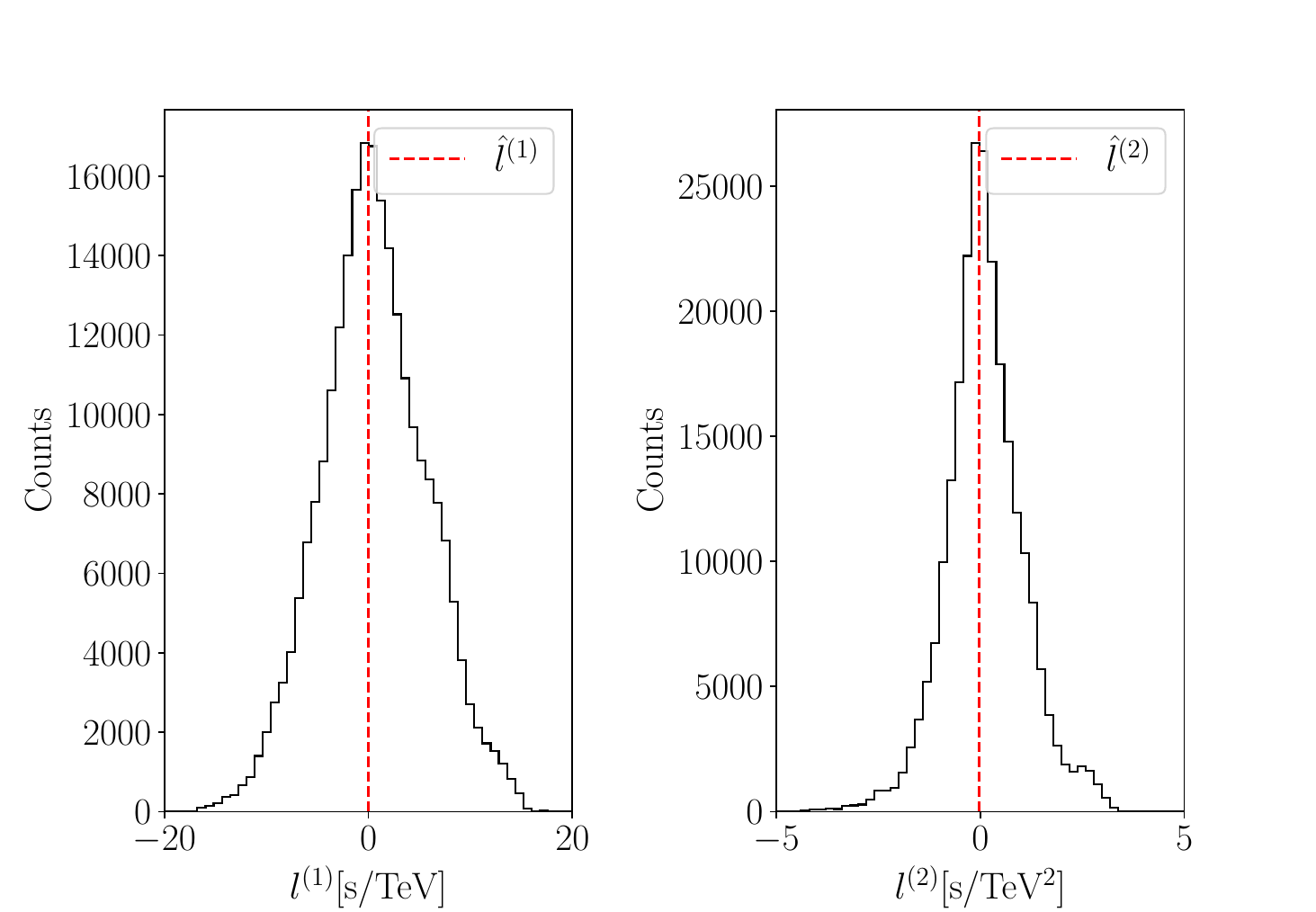}
    \caption{The distributions of $l^{(1)}$(left) and $l^{(2)}$(right). The red dashed lines represent the positions of the peaks identified using the kernel density estimation method.}
    \label{l_ij_distribution}
\end{figure}
\begin{figure}[htbp]
    \includegraphics[width=\linewidth]{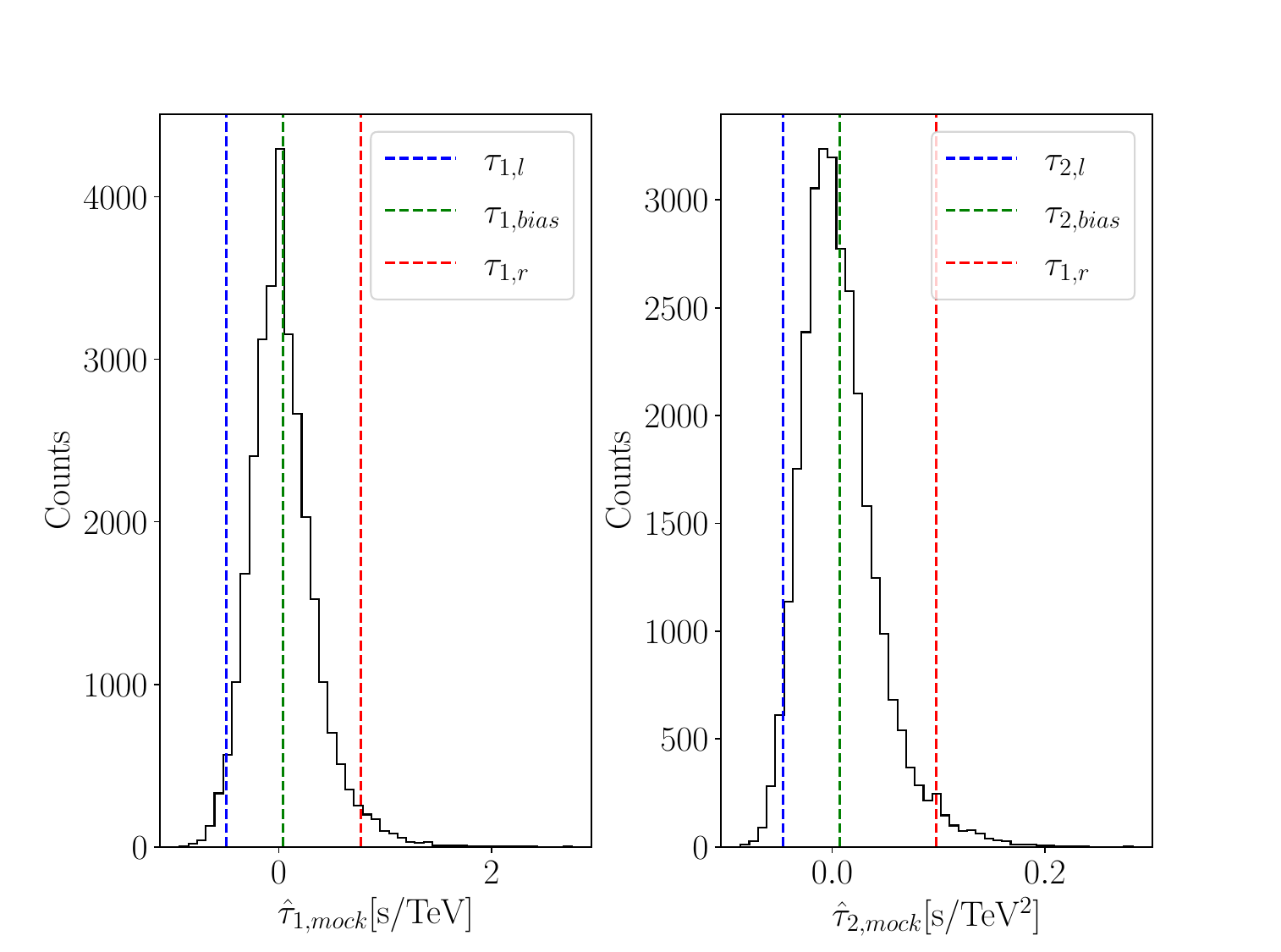}
    \caption{The distribution of the 30,000 peaks obtained from the 30,000 mock data sets by the PV analysis is illustrated in the left and right panels for the case $n=1$ and $n=2$, respectively. The green dashed lines represent the bias, while the blue and red dashed lines depict the $2.5\%$ and $97.5\%$ quantiles, respectively.}
    \label{peaksdistribution}
\end{figure}

The distributions of $l^{(1)}$ and $l^{(2)}$ are shown in the left and right panels of Fig.~\ref{l_ij_distribution}, respectively. In order to accurately pinpoint the peak position, we narrow our focus to spectral lags within the ranges of $ -5\,\text{s}/\text{TeV}\leq l^{(1)}_{i,j}\leq 5\,\text{s}/\text{TeV}$ for $n=1$ and $-1\,\text{s}/\text{TeV}^2\leq l^{(2)}_{i,j}\leq 1\,\text{s}/\text{TeV}^2$  for $n=2$. Subsequently, the peak positions are determined to be $\hat{\tau}_1=-0.045\,\text{s}/\text{TeV}$ and $\hat{\tau}_2=-0.019\,\text{s}/\text{TeV}^2$ for the respective cases, illustrated as red dashed lines in Fig.~\ref{l_ij_distribution}.

Subsequently, we proceed to establish the $95\%$ confidence intervals (CIs) for the LIV parameters using Monte Carlo simulation (for further details, see \cite{Vasileiou_2013}). As the estimator of the actual value $\tau_n$, $\hat{\tau}_n$ is expected to be distributed around $\tau_n$ according to a certain probability density function (PDF). We consider the PDF of $\hat{\tau}_n-\tau_n$, denoted as $P(\hat{\tau}_n-\tau_n)$, and assume it is independent of the actual value $\tau_n$. This allows us to take it as the PDF without LIV, denoted as $P(\hat{\tau}_{n,\text{mock}})$. $P(\hat{\tau}_{n,\text{mock}})$ can be obtained by generating mock data sets and conducting PV analysis in the absence of LIV. Specifically, we obtain $P(\hat{\tau}_{n,\text{mock}})$ through the following four steps:
\begin{itemize}
    \item In order to take into account the uncertainties of the  KM2A photon's estimated energies, we randomly sample their energies with Gaussian distributions, according to the mean values and uncertainties given by Ref. \cite{cao2023high}.
    \item In the absence of LIV, there is no correlation between the time and energy of photons. Therefore, we generate mock data by retaining the arrival time of all the 3,882 photons and randomly shuffling their energies.
    \item We repeat the above two steps to generate 30,000 sets of mock data.
    \item We conduct the PV analysis for each data set, and obtain 30,000 peaks from all the data sets. Then we perform histogram statistics on these peaks to get $P(\hat{\tau}_{n,\text{mock}})$ (up to a normalization).
\end{itemize}

The distributions of these peaks are illustrated in Fig.~\ref{peaksdistribution}. In the absence of LIV, these peaks are expected to be distributed with a mean value equal to zero. However, the histograms of these peaks, as shown in Fig.~\ref{peaksdistribution}, reveal that their mean value slightly deviate from zero, indicating the inherent bias of the PV method, denoted as $\tau_{n,\text{bias}}$. The blue and red dashed lines in the figure represent the 2.5$\%$ and $97.5\%$ quantiles at $\tau_{n,l}$ and $\tau_{n,r}$, respectively.  Under the assumption that $P(\hat{\tau}_n-\tau_n)$ is independent of the actual value $\tau_n$, we have $P(\hat{\tau}_n-\tau_n)=P(\hat{\tau}_{n,\text{mock}})$ . Consequently, the lower and upper bounds of the 95$\%$ CI are given by \cite{Vasileiou_2013}
\begin{equation}
\quad \tau_{n,\text{LL}}=\hat{\tau}_n-\tau_{n,r}, \;\;\tau_{n,\text{UL}}=\hat{\tau}_n-\tau_{n,l}.
\end{equation}
Then the constraints on $\tau_n$ can be converted to constraints on $\eta_n$.  Furthermore, $\eta^\text{PV}_\text{LL}$ and $\eta^\text{PV}_\text{UL}$ can be converted to the lower limits of QG scale at the 95$\%$ confidence level (CL), as listed in Tab.~\ref{PVconstraint}.

\begin{table}[htbp]
    \centering
    \caption{The values of  $\eta^\text{PV}_\text{LL}$and $\eta^\text{PV}_\text{UL}$ obtained by the PV method, along with the corresponding lower limits on the QG energy at the $95\%$ CL. }
    \begin{ruledtabular}
    \begin{tabular}{c|cc}
    &$\eta^\text{PV}_\text{LL}$&$\eta^\text{PV}_\text{UL}$\\
    \hline
    $\eta_1$&-0.139&0.076 \\
    $\eta_2$&-0.015&0.003\\
    \hline
    & $\mathcal{S}=-1$&$\mathcal{S}=+1$\\
    \hline
    $E_{\text{QG},1}[10^{19}\text{GeV}]$&8.8&16.1\\
    $E_{\text{QG},2}[10^{11}\text{GeV}]$&9.9&20.7\\
    \end{tabular}
    \end{ruledtabular}
    \label{PVconstraint}
\end{table}

We proceed to examine the impact of the energy uncertainties of WCDA photons on the limits. We investigate this impact by sampling the estimated energies of WCDA photons. As the specific distribution of the 3,818 photons' energies within the $N_\text{hit}$ bin is not publicly available, we assume a Gaussian distribution with a mean of $E_0=0.35$ TeV and a standard deviation of $E_0$. We randomly sample the estimated energies of these 3,818 photons according to this distribution, adjusting energies less than 0.1 TeV to 0.1 TeV, and construct 300 datasets. We find that the constraints are, on average, altered by approximately $+2.5\%$ ($-1.8\%$) for the subluminal (superluminal) scenario for $n=1$, and by $-1.5\%$ ($+2.8\%$) for the subluminal (superluminal) scenario for $n=2$.

Another factor that could potentially affect the results is the finite time bin of WCDA and KM2A. Similar to the analysis of energy uncertainties, we generate 300 datasets, with each dataset featuring randomly distributed
arrival times for every photon within its respective 0.1 s time bin. As a result, the arrival times of photons in a time bin vary across the 300 datasets. We find that this uncertainty leads to fluctuations of $5.6\%$($6.7\%$) in the constraint results for the subluminal (superluminal) scenario for $n=1$, and $4.3\%$($3.0\%$) for the subluminal (superluminal) secnario for $n=2$.

\section{Maximum Likelihood Analysis\label{MLM}}

\subsection{method\label{method}}

The other event by event method considered in this study is the ML method. This unbinned method was initially introduced to the study of LIV  by Mart\' {i}nez and Errando in \cite{Mart_nez_2009}. They applied this method to the
MAGIC observation of Mrk 501 and the H.E.S.S. observation of PKS 2155-304, and demonstrated its capacity to yield a more stronger constraint on $E_\text{QG}$ compared to other methods.  
The efficacy of the ML method is derived from its ability to incorporate all the physical information from the time-of-flight analysis, including the arrival time and estimated energy of each photon, based on the PDF.
However, the original ML method did not consider the impact of background. This was addressed in the analysis of PG 1553+113 due to the low signal-to-noise ratio \cite{Abramowski_2015}.  Given that only 142 photons are observed by KM2A in the time interval 230-900 s, with approximately 17.6 of them being cosmic ray events, it is necessary to consider the background in accordance with cosmic ray background estimation \cite{cao2023high}. 

The PDF of observing the $i$-th photon with the estimated energy $E^{(i)}_\text{est}$ at time $t^{(i)}$ is expressed as \cite{Acciari_2020} 
\begin{widetext}
\begin{equation}
P(t^{(i)},E^{(i)}_\text{est}\left|\eta_n\right.)=p_s\frac{f_s(t^{(i)},E^{(i)}_\text{est}\left|\eta_n\right.)}{\int_{E_\text{min}}^{E_\text{max}}dE_\text{est}\int_{t_\text{min}}^{t_\text{max}}dt f_s(t,E_\text{est}\left|\eta_n\right.)}+p_b\frac{f_b(t^{(i)},E^{(i)}_\text{est})}{\int_{E_\text{min}}^{E_\text{max}}dE_\text{est}\int_{t_\text{min}}^{t_\text{max}}dt f_b(t,E_\text{est})},
    \label{P}
\end{equation}
\end{widetext}
where $p_s=(N_\text{on}-N_\text{b})/N_\text{on}$ and $p_b=N_\text{b}/N_\text{on}$ are the probabilities of this photon being signal and background, respectively, $N_\text{on}$ and $N_\text{b}$ are total and background event numbers, respectively, $f_s(t,E_\text{est}\left|\eta_n\right.)$ denotes the PDF (up to a normalization) of a signal photon arriving at time $t$ with the estimated energy $E_\text{est}$, and $f_b$ denotes the PDF of the background. In the normalization factors, $\left[E_\text{min},E_\text{max}\right]=\left[3,13\right]$ TeV and  $[t_\text{min},t_\text{max}]=\left[230,900\right]$ s represent the ranges of the estimated energy and observed time being analyzed, respectively. For the KM2A observation, we take $N_\text{on}=64$ and $N_\text{b}=3.3$ in the interval 230-300 s, and $N_\text{on}=78$ and $N_\text{b}=14.3$ in the interval 300-900 s. Consequently, $p_s$ and $p_b$ have different values in the two time intervals. The background PDF $f_b$ is assumed to be independent of time within 230-300 s and 300-900 s.  Its dependence on energy is obtained by interpolating the estimated background events in different energy bins. 

The PDF function of the signal $f_s(t,E_\text{est}\left|\eta_n\right.)$ is given by
\begin{widetext}
\begin{equation}
    f_s(t,E_\text{est}\left|\eta_n\right.)\propto \int_0^\infty dE\lambda\left(t-\Delta t_\text{LIV}(\eta_n,E,0)\right)\frac{dN}{dE}\left(t-\Delta t_\text{LIV}(\eta_n,E,0),E\right)F(E)A^{\text{KM2A}}_\text{eff}(t,E)G(E_\text{est},E),
    \label{f_s}
\end{equation}
\end{widetext}
where $E$ is the actual energy of the photon. $\lambda(t)$ and $dN(t,E)/dE$ represent the intrinsic temporal profile and energy spectrum of the GRB photons, respectively, and $\lambda(t)$ is assumed to be independent of energy. 
 $G(E_\text{est},E)$ denotes the energy resolution of KM2A, assumed to be a Gaussian function of $E_\text{est}$, with a mean of $E$ and a standard deviation of $0.4E$ \cite{cao2023high}. $F(E)=e^{-\tau(E)}$ with the optical depth $\tau(E)$ accounts for the EBL attenuation effect. $A^\text{KM2A}_\text{eff}(t,E)$ represents the effective area of the KM2A detector. For simplicity,  we use the average value of the effective areas at 230 s and 900 s \cite{cao2023high}, and ignore its time dependence.

In the presence of LIV, photons emitted simultaneously would arrive with an energy-dependent time lag $\Delta t_\text{LIV}$, compared to the scenario without LIV. In this case, the profile $\lambda(t)$ cannot be precisely inferred from observations on the Earth. Many previous studies, such as \cite{Mart_nez_2009,Abramowski_2011,Vasileiou_2013,Abramowski_2015}, derive $\lambda(t)$ from the measured light curve of the photons with lower energies, and only utilize the photons with higher energies to calculate the likelihood function. This approach is valid when the impact of LIV on the photons with lower energies is much less significant compared with those with higher energies.
In this study, we also adopt this approach, and utilize $\lambda(t)$ derived from the WCDA photons in the lowest $N_\text{hit}$ bin of $N_\text{hit}\in [30,33)$ \cite{2023}. In other words, the $\lambda(t)$ we employ is obtained by fitting the light curve of the lowest energy portion of WCDA photons, and is expressed as:
\begin{widetext}
\begin{equation}
    \lambda(t)=\left\{\begin{array}{ll}
        0 &,\; t<226\\
        \left[\left(\frac{t_{b,0}}{t_{b,1}}\right)^{-\omega_1\alpha_1}+\left(\frac{t_{b,0}}{t_{b,1}}\right)^{-\omega_1\alpha_2}\right]^{-1/\omega_1}\left(\frac{t-226}{t_{b,0}}\right)^{\alpha_0}&,\; 226\leq t< t_{b,0}+226\\
        \left[\left(\frac{t-226}{t_{b,1}}\right)^{-\omega_1\alpha_1}+\left(\frac{t-226}{t_{b,1}}\right)^{-\omega_1\alpha_2}\right]^{-1/\omega_1}&,\; t_{b,0}+226\leq t< t_{b,2}+226\\
        \left[\left(\frac{t_{b,2}}{t_{b,1}}\right)^{-\omega_1\alpha_1}+\left(\frac{t_{b,2}}{t_{b,1}}\right)^{-\omega_1\alpha_2}\right]^{-1/\omega_1}\left(\frac{t-226}{t_{b,2}}\right)^{\alpha_3}&,\; t\geq t_{b,2}+226.\\
    \end{array}\right.        
\end{equation}
\end{widetext}
In this section, we choose the values of the parameters $\left\{\omega_1,\alpha_0, \alpha_1, \alpha_2,\alpha_3, t_{b,0}, t_{b,1}, t_{b,2}\right\}$ to be the best-fit values $I_b=\left\{1.46,12.8,1.7,-1.108,-1.87,4.85,16.1,560\right\}$ \cite{2023} . Here, we assume a sharp transition from the slow decay phase to the steep decay phase with $\omega_2\to\infty$. The impact of uncertainties in these parameters is discussed in Appendix \ref{SE}.
Similar to the previous PV analysis, we continue to assume that the energies of all the WCDA photons are equivalent to the median energy  $E_0=0.35$ TeV. Subsequently, $\Delta t_\text{LIV}(\eta_n,E,0)$ in Eq.~\ref{f_s} is substituted  with the time lag between the photons with energies of $E$ and $E_0$ as $\Delta t_\text{LIV}(\eta_n,E,E_0)$.

In this work, we assume that the intrinsic temporal distribution $\lambda(t)$ is independent of energy, while consider the time dependence of the energy spectrum. The propagation of photons from the source to the Earth involves interactions with the EBL photons, leading to the production of electron-positron pairs. Consequently, a fraction of photons is absorbed and fails to reach the Earth. This effect represented by $F(E)$ depends on the photon energy. However, the EBL model itself is subject to considerable uncertainties. In this context, three distinct models are under consideration for the intrinsic energy spectrum of the photons $dN(t, E)/dE$ and the EBL absorption term $F(E)$:
\begin{itemize}
\item Model-1: power-law spectrum with a spectral index evolving continuously with time  $(E/\text{TeV})^{-2.579+0.135\lg(t-226)} $ \cite{2023}, and the EBL model of \cite{Saldana_Lopez_2021}. 
\item Model-2: power-law spectrum with a spectral index of -2.35 in 230-300 s and -2.26 in 300-900 s \cite{cao2023high}, and the EBL model of \cite{Saldana_Lopez_2021}.
\item Model-3: power-law spectrum with a spectral index of -2.12 in 230-300 s and log-parabolic spectrum $(E/\text{TeV})^{-2.03-0.15\log(E/\text{TeV})} $ in 300-900 s \cite{cao2023high}, and the LHAASO constrained EBL model \cite{cao2023high}. 
\end{itemize}

Model-1 is derived from fitting the WCDA observational data for the energy spectrum at 15 different time intervals.  Model-2 considers both WCDA and KM2A data for the energy spectrum, but it falls below experimental data points at the high-energy end. In order to achieve a more accurate fit, Model-3 is adjusted to better match the high-energy end by suppressing the EBL absorption. 

It is important to emphasize that all these available models are derived from the experimental data without including the LIV effects in the analysis. However, LIV effects can introduce discrepancies between the inferred spectra and the true spectra. For instance, in the subluminal scenario, where high-energy photons have slower speeds, the fraction of observed high-energy photons in the total photons increases with time. This results in a harder inferred spectrum compared to the true spectrum. Moreover, LIV effects can alter the threshold energies of interaction processes for photons, and effectively affect the EBL absorption effect.
Nevertheless, we will discuss the impact of these effects in Sec.~\ref{RD}, and argue that they can be safely negligible, thereby affirming the reliability of the aforementioned spectra. 
\begin{figure}[htbp]
    \centering
    \includegraphics[width=\linewidth]{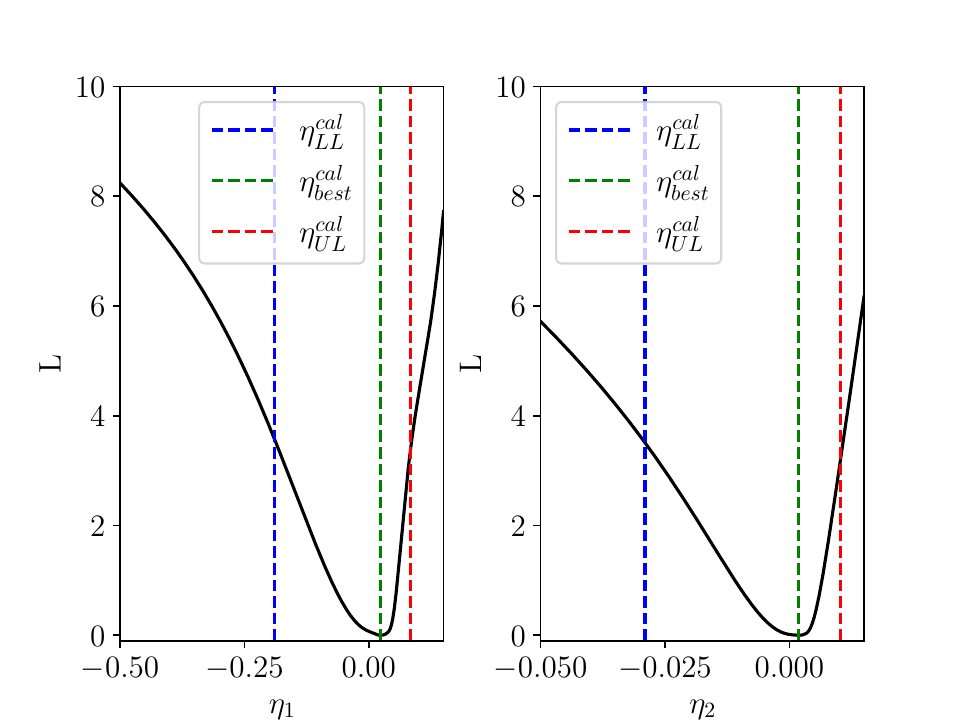}
    \caption{The $L(\eta_1)$ (left) and $L(\eta_2)$ (right) curves (black solid lines) of Model-1. The green dashed lines represent the position of $\eta^\text{cal}_\text{best}$. The red dashed lines and blue dashed lines represent the the lower and upper ranges of the 95$\%$ CI, respectively.}
    \label{M1}
\end{figure}
\begin{figure}[htbp]
    \centering
    \includegraphics[width=\linewidth]{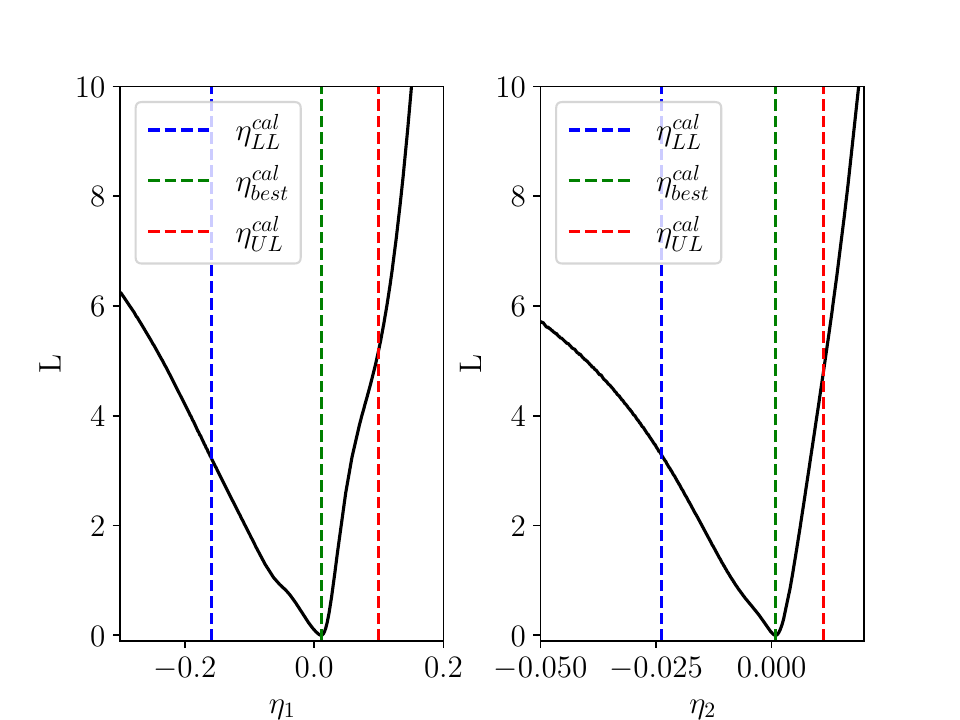}
    \caption{Same as Fig.~\ref{M1}, but for Model-2.}
    \label{M2}
\end{figure}

\begin{figure}[htbp]
    \centering
    \includegraphics[width=\linewidth]{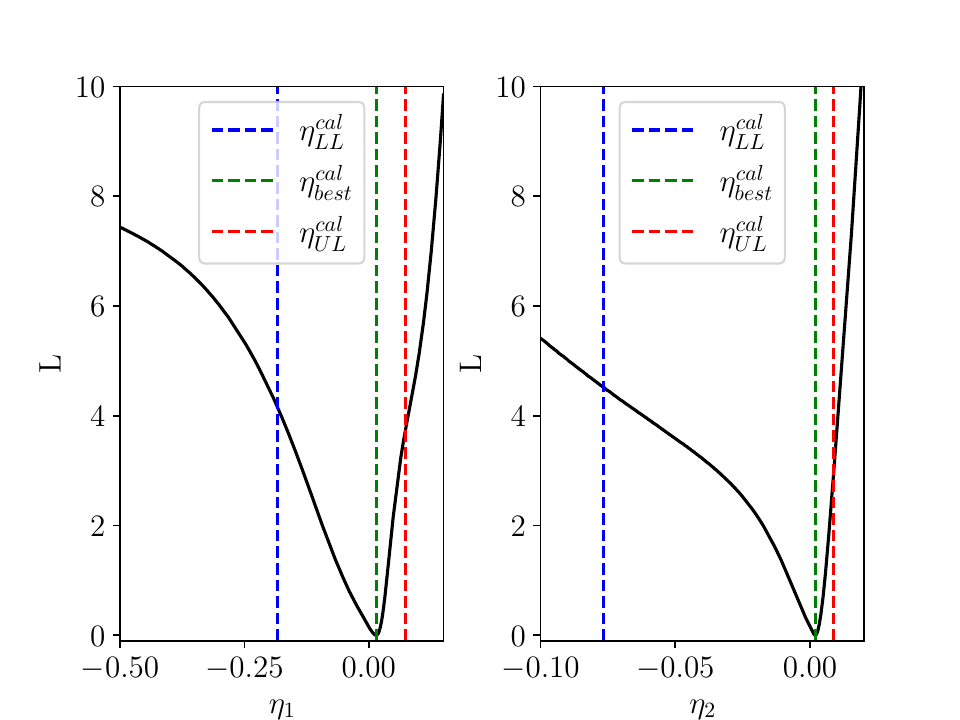}
    \caption{Same as Fig.~\ref{M1}, but for Model-3.}
    \label{M3}
\end{figure}

The likelihood function of the 142 KM2A photons can be expressed as
\begin{equation}
    \mathcal{L}\left(\eta_n\left|\left\{t^{(i)},E^{(i)}_\text{est}\right\}\right.\right)=\prod_{i=1}^{142}P(t^{(i)},E^{(i)}_\text{est}\left|\eta_n\right.).
    \label{mathL}
\end{equation}
The best value $\hat{\eta}_n$ is determined by adjusting $\eta_n$ to maximize the above likelihood function. We then define
\begin{equation}
    L(\eta_n)\equiv -2\ln\left(\frac{\mathcal{L}_{\eta_n}}{\mathcal{L}_{\hat{\eta}_n}}\right),
\end{equation}
where $\mathcal{L}_{\eta_n}$ is given by Eq.~\ref{mathL}, and $L_{\hat{\eta}_n}$ is the value of the likelihood when $\eta_n=\hat{\eta}_n$. We use Monte Carlo simulations to construct the calibrated CIs of $\eta_n$ (refer to Appendix \ref{CI}). The lower and upper limit are determined by two thresholds $L_l$ and $L_r$ on the left and right sides of the minimum point of the $L(\eta_n)$ curve, respectively. The constraints obtained with the ML method for three models are depicted in Fig.~\ref{M1}, Fig.~\ref{M2}, Fig.~\ref{M3}, and Tab.~\ref{Tab2}.

\begin{table*}[htbp]
    \centering
    \caption{The calibrated best fits of $\eta_n$, the lower and upper bounds of the 95$\%$ CIs of $\eta_n$, and the corresponding 95$\%$ CL lower limits on the $E_\text{QG}$.}
    \begin{ruledtabular}
    \begin{tabular}{c|ccc|ccc|cc|cc}
    &\multicolumn{3}{c|}{$\eta_1$}&\multicolumn{3}{c|}{$\eta_2$}&\multicolumn{2}{c|}{$E_\text{QG,1}\left[10^{19}\text{GeV}\right]$}&\multicolumn{2}{c}{$E_\text{QG,2}\left[10^{11}\text{GeV}\right]$}\\
    &$\eta^\text{cal}_\text{LL}$& $\eta^\text{cal}_\text{best}$&$\eta^\text{cal}_\text{UL}$&$\eta^\text{cal}_\text{LL}$& $\eta^\text{cal}_\text{best}$&$\eta^\text{cal}_\text{UL}$& $\mathcal{S}=+1$&$\mathcal{S}=-1$&$\mathcal{S}=+1$&$\mathcal{S}=-1$\\
    \hline
    Model-1& -0.189&0.024&0.083&-0.029&0.002
    &0.010&14.7&6.5& 12.0&7.2\\
    Model-2& -0.159&0.011&0.100 &-0.024&0.001&0.011&12.2&7.7&11.5&7.9\\
    Model-3& -0.184&0.015 &0.074&-0.077&0.002&0.009&16.4&6.6&13.1&4.4\\
    \end{tabular}
    \end{ruledtabular}
    \label{Tab2}
    \end{table*}

\subsection{Discussions\label{RD}}

Comparing Tab.~\ref{PVconstraint} and Tab.~\ref{Tab2}, it is evident that the constraints obtained through the PV method are at the same order of  those obtained through the ML method, with the PV method yielding slightly stronger constraints. Additionally, the three different spectral and EBL models in Tab.~\ref{Tab2} also yield similar constraint results.

A notable feature of the results from both methods is that the constraints on the superluminal case are considerably weaker than those on the subluminal case. This discrepancy is particularly pronounced in the ML analysis, as evidenced by the markedly slower ascent in the negative $\eta_n$ direction in comparison to the positive $\eta_n$ direction, depicted in the $L(\eta_n)$ curves in Fig.~\ref{M1}, Fig.~\ref{M2}, and Fig.~\ref{M3}. Similar to the discussion of the minimal model in \cite{Acciari_2020}, this feature can be attributed to the rapid rise and slow decay of the light curve. In the subluminal scenario, photons of higher energies have slower velocities, and are anticipated to originate from an earlier time within the temporal profile.  Along with the decreasing direction of time, the temporal profile sharply decreases before the transition. This phenomenon reduces the contributions of many high energy photons to the likelihood function, and results in a rapid increase of $L(\eta_n)$ with increasing $\eta_n$. Conversely, in the superluminal scenario, higher energy photons have faster velocities and are expected to originate from a later time within the temporal profile. These photons may not significantly impact the likelihood within the time span, where the afterglow has a slow decay and long duration. 
Consequently, $L(\eta_n)$ exhibits a slow increase in the negatively $\eta_n$ direction. 

We conduct a validation using  data within 230-300 s and altering the temporal profile to sharply decrease after 300 s.  
We find that $L(\eta_n)$ also sharply increases in the negative direction of $\eta_n$ in this case. This is consistent with the above discussions. In summary, our analysis indicates that the GRB afterglow with the characteristic of rapid rise and slow decay is more sensitive to the subluminal LIV effect, and can provide stronger constraints compared to the superluminal case.

Then we discuss the rationality of the energy spectrum used in the ML analysis as follows. For Model-1, in Ref.~\cite{cao2023high}, a time-hardening spectrum is obtained by fitting the WCDA results in 15 time intervals, with the shortest time interval being approximately 4 s. If this time-hardening behavior is not an intrinsic property of GRB 221009A but rather caused by LIV effects, it should be caused by the subluminal LIV effect. However, we have imposed strict constraints on the subluminal case. The energy range of the WCDA photons is approximately 0.2-7 TeV. According to the results of the PV analysis, given a most energetic photon with 7 TeV, the maximum time delays caused by subluminal LIV are approximately 3 s and 1 s for $n=1$ and $n=2$ cases, respectively. These time delays are less than 4 s, which is the shortest time interval to derive the time dependent energy spectrum. Moreover, only a small fraction of the photons is at the high-energy end. This means that the time delays in the energy spectra  caused by LIV for the majority of the photons should be much less than 4 s, and can not induce the observed time-hardening behaviour of the energy  spectra. Additionally, the power-law spectrum in 300-900 s is also harder than that in 230-300 s in Model-2. These time intervals are much longer than that in Model-1, which further indicates that the time-hardening can not be attributed to LIV, but rather represents an intrinsic property of GRB 221009A.

For EBL absorption, only the subluminal case raises the threshold energy of the pair production, thereby reducing EBL absorption. For the $n=1$ case, according to the constraints set by \cite{cao2023high}, when $E_{\text{QG},1}>2.8 M_{PL}$, there is no significant difference in $F(E)$ between the cases with and without LIV for photons with energies $\lesssim 10$ TeV. The constraint on $E_{\text{QG},1}$ obtained in our analysis is much tighter than $2.8 M_{PL}$ in the subluminal scenario, indicating that LIV cannot cause a significant weakening of EBL absorption within the energy range of KM2A photons considered here.

We note that a previous study \cite{piran2023lorentz} also utilized LHAASO data to establish constraints on the LIV energy scale. At the time of that study, only the data from WCDA had been published, and the data from KM2A were not available. The authors of \cite{piran2023lorentz} employed a method that involved fitting the light curve and spectrum and minimizing the $\chi^2$ value to obtain the confidence interval of the LIV parameters. In contrast, we utilize data from both WCDA and KM2A in this study. In order to fully exploit the information carried by the observation, we employ two distinct event by event methods, including the PV and ML methods. In terms of the results, our constraints on the superluminal scenario are comparable to those in \cite{piran2023lorentz}, whereas the constraints on the subluminal scenario that we obtain are stronger than those in \cite{piran2023lorentz}.

 \section{Conclusions\label{Sum}}
In this study, we employ two independent methods, namely the PV method and the ML method, to investigate the constraints on the QG energy scale from observations of GRB 221009A by LHAASO. Similar constraints are obtained using these two methods. Specifically, using the PV method, the $95\%$ CL lower limits are determined to be $E_{\text{QG},1} > 16.1\, (8.8)\times 10^{19}$GeV  for the subluminal (superluminal) scenario of $n=1$, and $E_{\text{QG},2} > 20.7 \,(9.9) \times 10^{11}$GeV for the subluminal (superluminal) scenario of $n=2$. Through the ML method, the $95\%$ CL lower limits are $E_{\text{QG},1} > 14.7\, (6.5)\times 10^{19}$GeV  for the subluminal (superluminal) scenario of $n=1$, and $E_{\text{QG},2} > 12.0 \,(7.2) \times 10^{11}$GeV for the subluminal (superluminal) scenario of $n=2$.  It is observed that the extended duration of the GRB afterglow decay makes it less sensitive to the superluminal scenario. On the other hand, the rapid rise phase makes it more suitable for studying the subluminal scenario, leading to stronger constraints. We also discuss the impact of the intrinsic uncertainties in the light curve fitting on the constraint results, and find that the constraints are altered by several tens of percent.
\acknowledgments
This work is supported by the National Natural Science Foundation of China under grant No. 12175248.


\bibliography{LIV}

\begin{thebibliography}{50}%
\makeatletter
\providecommand \@ifxundefined [1]{%
 \@ifx{#1\undefined}
}%
\providecommand \@ifnum [1]{%
 \ifnum #1\expandafter \@firstoftwo
 \else \expandafter \@secondoftwo
 \fi
}%
\providecommand \@ifx [1]{%
 \ifx #1\expandafter \@firstoftwo
 \else \expandafter \@secondoftwo
 \fi
}%
\providecommand \natexlab [1]{#1}%
\providecommand \enquote  [1]{``#1''}%
\providecommand \bibnamefont  [1]{#1}%
\providecommand \bibfnamefont [1]{#1}%
\providecommand \citenamefont [1]{#1}%
\providecommand \href@noop [0]{\@secondoftwo}%
\providecommand \href [0]{\begingroup \@sanitize@url \@href}%
\providecommand \@href[1]{\@@startlink{#1}\@@href}%
\providecommand \@@href[1]{\endgroup#1\@@endlink}%
\providecommand \@sanitize@url [0]{\catcode `\\12\catcode `\$12\catcode `\&12\catcode `\#12\catcode `\^12\catcode `\_12\catcode `\%12\relax}%
\providecommand \@@startlink[1]{}%
\providecommand \@@endlink[0]{}%
\providecommand \url  [0]{\begingroup\@sanitize@url \@url }%
\providecommand \@url [1]{\endgroup\@href {#1}{\urlprefix }}%
\providecommand \urlprefix  [0]{URL }%
\providecommand \Eprint [0]{\href }%
\providecommand \doibase [0]{https://doi.org/}%
\providecommand \selectlanguage [0]{\@gobble}%
\providecommand \bibinfo  [0]{\@secondoftwo}%
\providecommand \bibfield  [0]{\@secondoftwo}%
\providecommand \translation [1]{[#1]}%
\providecommand \BibitemOpen [0]{}%
\providecommand \bibitemStop [0]{}%
\providecommand \bibitemNoStop [0]{.\EOS\space}%
\providecommand \EOS [0]{\spacefactor3000\relax}%
\providecommand \BibitemShut  [1]{\csname bibitem#1\endcsname}%
\let\auto@bib@innerbib\@empty
\bibitem [{\citenamefont {Burgess}(2004)}]{Burgess_2004}%
  \BibitemOpen
  \bibfield  {author} {\bibinfo {author} {\bibfnamefont {C.~P.}\ \bibnamefont {Burgess}},\ }\bibfield  {title} {\bibinfo {title} {{Quantum Gravity in Everyday Life: General Relativity as an Effective Field Theory}},\ }\bibfield  {journal} {\bibinfo  {journal} {Living Reviews in Relativity}\ }\textbf {\bibinfo {volume} {7}},\ \href {https://doi.org/10.12942/lrr-2004-5} {10.12942/lrr-2004-5} (\bibinfo {year} {2004})\BibitemShut {NoStop}%
\bibitem [{\citenamefont {Deser}\ and\ \citenamefont {van Nieuwenhuizen}(1974)}]{PhysRevD.10.411}%
  \BibitemOpen
  \bibfield  {author} {\bibinfo {author} {\bibfnamefont {S.}~\bibnamefont {Deser}}\ and\ \bibinfo {author} {\bibfnamefont {P.}~\bibnamefont {van Nieuwenhuizen}},\ }\bibfield  {title} {\bibinfo {title} {{Nonrenormalizability of the quantized Dirac-Einstein system}},\ }\href {https://doi.org/10.1103/PhysRevD.10.411} {\bibfield  {journal} {\bibinfo  {journal} {Phys. Rev. D}\ }\textbf {\bibinfo {volume} {10}},\ \bibinfo {pages} {411} (\bibinfo {year} {1974})}\BibitemShut {NoStop}%
\bibitem [{\citenamefont {Addazi}\ \emph {et~al.}(2022)\citenamefont {Addazi} \emph {et~al.}}]{Addazi_2022}%
  \BibitemOpen
  \bibfield  {author} {\bibinfo {author} {\bibfnamefont {A.}~\bibnamefont {Addazi}} \emph {et~al.},\ }\bibfield  {title} {\bibinfo {title} {{Quantum gravity phenomenology at the dawn of the multi-messenger era{\textemdash}A review}},\ }\href {https://doi.org/10.1016/j.ppnp.2022.103948} {\bibfield  {journal} {\bibinfo  {journal} {Progress in Particle and Nuclear Physics}\ }\textbf {\bibinfo {volume} {125}},\ \bibinfo {pages} {103948} (\bibinfo {year} {2022})}\BibitemShut {NoStop}%
\bibitem [{\citenamefont {Bonanno}\ and\ \citenamefont {Reuter}(1999)}]{Bonanno_1999}%
  \BibitemOpen
  \bibfield  {author} {\bibinfo {author} {\bibfnamefont {A.}~\bibnamefont {Bonanno}}\ and\ \bibinfo {author} {\bibfnamefont {M.}~\bibnamefont {Reuter}},\ }\bibfield  {title} {\bibinfo {title} {{Quantum gravity effects near the null black hole singularity}},\ }\bibfield  {journal} {\bibinfo  {journal} {Physical Review D}\ }\textbf {\bibinfo {volume} {60}},\ \href {https://doi.org/10.1103/physrevd.60.084011} {10.1103/physrevd.60.084011} (\bibinfo {year} {1999})\BibitemShut {NoStop}%
\bibitem [{\citenamefont {Bosma}\ \emph {et~al.}(2019)\citenamefont {Bosma}, \citenamefont {Knorr},\ and\ \citenamefont {Saueressig}}]{Bosma_2019}%
  \BibitemOpen
  \bibfield  {author} {\bibinfo {author} {\bibfnamefont {L.}~\bibnamefont {Bosma}}, \bibinfo {author} {\bibfnamefont {B.}~\bibnamefont {Knorr}},\ and\ \bibinfo {author} {\bibfnamefont {F.}~\bibnamefont {Saueressig}},\ }\bibfield  {title} {\bibinfo {title} {{Resolving Spacetime Singularities within Asymptotic Safety}},\ }\bibfield  {journal} {\bibinfo  {journal} {Physical Review Letters}\ }\textbf {\bibinfo {volume} {123}},\ \href {https://doi.org/10.1103/physrevlett.123.101301} {10.1103/physrevlett.123.101301} (\bibinfo {year} {2019})\BibitemShut {NoStop}%
\bibitem [{\citenamefont {Gambini}\ \emph {et~al.}(2014)\citenamefont {Gambini}, \citenamefont {Olmedo},\ and\ \citenamefont {Pullin}}]{Gambini_2014}%
  \BibitemOpen
  \bibfield  {author} {\bibinfo {author} {\bibfnamefont {R.}~\bibnamefont {Gambini}}, \bibinfo {author} {\bibfnamefont {J.}~\bibnamefont {Olmedo}},\ and\ \bibinfo {author} {\bibfnamefont {J.}~\bibnamefont {Pullin}},\ }\bibfield  {title} {\bibinfo {title} {Quantum black holes in loop quantum gravity},\ }\href {https://doi.org/10.1088/0264-9381/31/9/095009} {\bibfield  {journal} {\bibinfo  {journal} {Classical and Quantum Gravity}\ }\textbf {\bibinfo {volume} {31}},\ \bibinfo {pages} {095009} (\bibinfo {year} {2014})}\BibitemShut {NoStop}%
\bibitem [{\citenamefont {Amelino-Camelia}(2013)}]{Amelino_Camelia_2013}%
  \BibitemOpen
  \bibfield  {author} {\bibinfo {author} {\bibfnamefont {G.}~\bibnamefont {Amelino-Camelia}},\ }\bibfield  {title} {\bibinfo {title} {{Quantum-Spacetime Phenomenology}},\ }\bibfield  {journal} {\bibinfo  {journal} {Living Reviews in Relativity}\ }\textbf {\bibinfo {volume} {16}},\ \href {https://doi.org/10.12942/lrr-2013-5} {10.12942/lrr-2013-5} (\bibinfo {year} {2013})\BibitemShut {NoStop}%
\bibitem [{\citenamefont {Kosteleck\'y}\ and\ \citenamefont {Samuel}(1989)}]{PhysRevD.39.683}%
  \BibitemOpen
  \bibfield  {author} {\bibinfo {author} {\bibfnamefont {V.~A.}\ \bibnamefont {Kosteleck\'y}}\ and\ \bibinfo {author} {\bibfnamefont {S.}~\bibnamefont {Samuel}},\ }\bibfield  {title} {\bibinfo {title} {{Spontaneous breaking of Lorentz symmetry in string theory}},\ }\href {https://doi.org/10.1103/PhysRevD.39.683} {\bibfield  {journal} {\bibinfo  {journal} {Phys. Rev. D}\ }\textbf {\bibinfo {volume} {39}},\ \bibinfo {pages} {683} (\bibinfo {year} {1989})}\BibitemShut {NoStop}%
\bibitem [{\citenamefont {Carroll}\ \emph {et~al.}(2001)\citenamefont {Carroll}, \citenamefont {Harvey}, \citenamefont {Kosteleck{\'{y} }}, \citenamefont {Lane},\ and\ \citenamefont {Okamoto}}]{Carroll_2001}%
  \BibitemOpen
  \bibfield  {author} {\bibinfo {author} {\bibfnamefont {S.~M.}\ \bibnamefont {Carroll}}, \bibinfo {author} {\bibfnamefont {J.~A.}\ \bibnamefont {Harvey}}, \bibinfo {author} {\bibfnamefont {V.~A.}\ \bibnamefont {Kosteleck{\'{y} }}}, \bibinfo {author} {\bibfnamefont {C.~D.}\ \bibnamefont {Lane}},\ and\ \bibinfo {author} {\bibfnamefont {T.}~\bibnamefont {Okamoto}},\ }\bibfield  {title} {\bibinfo {title} {{Noncommutative Field Theory and Lorentz Violation}},\ }\bibfield  {journal} {\bibinfo  {journal} {Physical Review Letters}\ }\textbf {\bibinfo {volume} {87}},\ \href {https://doi.org/10.1103/physrevlett.87.141601} {10.1103/physrevlett.87.141601} (\bibinfo {year} {2001})\BibitemShut {NoStop}%
\bibitem [{\citenamefont {Becker}\ \emph {et~al.}(2006)\citenamefont {Becker}, \citenamefont {Becker},\ and\ \citenamefont {Schwarz}}]{becker2006string}%
  \BibitemOpen
  \bibfield  {author} {\bibinfo {author} {\bibfnamefont {K.}~\bibnamefont {Becker}}, \bibinfo {author} {\bibfnamefont {M.}~\bibnamefont {Becker}},\ and\ \bibinfo {author} {\bibfnamefont {J.~H.}\ \bibnamefont {Schwarz}},\ }\href@noop {} {\emph {\bibinfo {title} {{String theory and M-theory: A modern introduction}}}}\ (\bibinfo  {publisher} {Cambridge university press},\ \bibinfo {year} {2006})\BibitemShut {NoStop}%
\bibitem [{\citenamefont {Mukhi}(2011)}]{Mukhi_2011}%
  \BibitemOpen
  \bibfield  {author} {\bibinfo {author} {\bibfnamefont {S.}~\bibnamefont {Mukhi}},\ }\bibfield  {title} {\bibinfo {title} {String theory: a perspective over the last 25 years},\ }\href {https://doi.org/10.1088/0264-9381/28/15/153001} {\bibfield  {journal} {\bibinfo  {journal} {Classical and Quantum Gravity}\ }\textbf {\bibinfo {volume} {28}},\ \bibinfo {pages} {153001} (\bibinfo {year} {2011})}\BibitemShut {NoStop}%
\bibitem [{\citenamefont {Mattingly}(2005)}]{Mattingly_2005}%
  \BibitemOpen
  \bibfield  {author} {\bibinfo {author} {\bibfnamefont {D.}~\bibnamefont {Mattingly}},\ }\bibfield  {title} {\bibinfo {title} {{Modern Tests of Lorentz Invariance}},\ }\bibfield  {journal} {\bibinfo  {journal} {Living Reviews in Relativity}\ }\textbf {\bibinfo {volume} {8}},\ \href {https://doi.org/10.12942/lrr-2005-5} {10.12942/lrr-2005-5} (\bibinfo {year} {2005})\BibitemShut {NoStop}%
\bibitem [{\citenamefont {Reuter}(1998)}]{PhysRevD.57.971}%
  \BibitemOpen
  \bibfield  {author} {\bibinfo {author} {\bibfnamefont {M.}~\bibnamefont {Reuter}},\ }\bibfield  {title} {\bibinfo {title} {Nonperturbative evolution equation for quantum gravity},\ }\href {https://doi.org/10.1103/PhysRevD.57.971} {\bibfield  {journal} {\bibinfo  {journal} {Phys. Rev. D}\ }\textbf {\bibinfo {volume} {57}},\ \bibinfo {pages} {971} (\bibinfo {year} {1998})}\BibitemShut {NoStop}%
\bibitem [{\citenamefont {Percacci}(2017)}]{percacci2017introduction}%
  \BibitemOpen
  \bibfield  {author} {\bibinfo {author} {\bibfnamefont {R.}~\bibnamefont {Percacci}},\ }\href@noop {} {\emph {\bibinfo {title} {An introduction to covariant quantum gravity and asymptotic safety}}},\ Vol.~\bibinfo {volume} {3}\ (\bibinfo  {publisher} {World Scientific},\ \bibinfo {year} {2017})\BibitemShut {NoStop}%
\bibitem [{\citenamefont {Bonanno}\ \emph {et~al.}(2020)\citenamefont {Bonanno}, \citenamefont {Eichhorn}, \citenamefont {Gies}, \citenamefont {Pawlowski}, \citenamefont {Percacci}, \citenamefont {Reuter}, \citenamefont {Saueressig},\ and\ \citenamefont {Vacca}}]{bonanno2020critical}%
  \BibitemOpen
  \bibfield  {author} {\bibinfo {author} {\bibfnamefont {A.}~\bibnamefont {Bonanno}}, \bibinfo {author} {\bibfnamefont {A.}~\bibnamefont {Eichhorn}}, \bibinfo {author} {\bibfnamefont {H.}~\bibnamefont {Gies}}, \bibinfo {author} {\bibfnamefont {J.~M.}\ \bibnamefont {Pawlowski}}, \bibinfo {author} {\bibfnamefont {R.}~\bibnamefont {Percacci}}, \bibinfo {author} {\bibfnamefont {M.}~\bibnamefont {Reuter}}, \bibinfo {author} {\bibfnamefont {F.}~\bibnamefont {Saueressig}},\ and\ \bibinfo {author} {\bibfnamefont {G.~P.}\ \bibnamefont {Vacca}},\ }\href@noop {} {\bibinfo {title} {Critical reflections on asymptotically safe gravity}} (\bibinfo {year} {2020}),\ \Eprint {https://arxiv.org/abs/2004.06810} {arXiv:2004.06810 [gr-qc]} \BibitemShut {NoStop}%
\bibitem [{\citenamefont {Colladay}\ and\ \citenamefont {Kosteleck{\'{y} }}(1997)}]{Colladay_1997}%
  \BibitemOpen
  \bibfield  {author} {\bibinfo {author} {\bibfnamefont {D.}~\bibnamefont {Colladay}}\ and\ \bibinfo {author} {\bibfnamefont {V.~A.}\ \bibnamefont {Kosteleck{\'{y} }}},\ }\bibfield  {title} {\bibinfo {title} {{CPT} violation and the standard model},\ }\href {https://doi.org/10.1103/physrevd.55.6760} {\bibfield  {journal} {\bibinfo  {journal} {Physical Review D}\ }\textbf {\bibinfo {volume} {55}},\ \bibinfo {pages} {6760} (\bibinfo {year} {1997})}\BibitemShut {NoStop}%
\bibitem [{\citenamefont {Kosteleck{\'{y} }}\ and\ \citenamefont {Mewes}(2009)}]{Kosteleck__2009}%
  \BibitemOpen
  \bibfield  {author} {\bibinfo {author} {\bibfnamefont {V.~A.}\ \bibnamefont {Kosteleck{\'{y} }}}\ and\ \bibinfo {author} {\bibfnamefont {M.}~\bibnamefont {Mewes}},\ }\bibfield  {title} {\bibinfo {title} {{Electrodynamics with Lorentz-violating operators of arbitrary dimension}},\ }\bibfield  {journal} {\bibinfo  {journal} {Physical Review D}\ }\textbf {\bibinfo {volume} {80}},\ \href {https://doi.org/10.1103/physrevd.80.015020} {10.1103/physrevd.80.015020} (\bibinfo {year} {2009})\BibitemShut {NoStop}%
\bibitem [{\citenamefont {Amelino-Camelia}(2002)}]{AMELINO_CAMELIA_2002}%
  \BibitemOpen
  \bibfield  {author} {\bibinfo {author} {\bibfnamefont {G.}~\bibnamefont {Amelino-Camelia}},\ }\bibfield  {title} {\bibinfo {title} {{Relativity} {in} {spacetimes} {with} {short}-{distance} {structure} {governed} {by} {an} {observer}-{independent} ({Planckian}) {length} {scale}},\ }\href {https://doi.org/10.1142/s0218271802001330} {\bibfield  {journal} {\bibinfo  {journal} {International Journal of Modern Physics D}\ }\textbf {\bibinfo {volume} {11}},\ \bibinfo {pages} {35} (\bibinfo {year} {2002})}\BibitemShut {NoStop}%
\bibitem [{\citenamefont {Kowalski-Glikman}()}]{Kowalski_Glikman}%
  \BibitemOpen
  \bibfield  {author} {\bibinfo {author} {\bibfnamefont {J.}~\bibnamefont {Kowalski-Glikman}},\ }\bibfield  {title} {\bibinfo {title} {{Introduction to Doubly Special Relativity}},\ }in\ \href {https://doi.org/10.1007/11377306_5} {\emph {\bibinfo {booktitle} {Planck Scale Effects in Astrophysics and Cosmology}}}\ (\bibinfo  {publisher} {Springer-Verlag})\ pp.\ \bibinfo {pages} {131--159}\BibitemShut {NoStop}%
\bibitem [{\citenamefont {{Biller}}\ \emph {et~al.}(1999)\citenamefont {{Biller}} \emph {et~al.}}]{1999PhRvL..83.2108B}%
  \BibitemOpen
  \bibfield  {author} {\bibinfo {author} {\bibfnamefont {S.~D.}\ \bibnamefont {{Biller}}} \emph {et~al.},\ }\bibfield  {title} {\bibinfo {title} {{Limits to Quantum Gravity Effects on Energy Dependence of the Speed of Light from Observations of TeV Flares in Active Galaxies}},\ }\href {https://doi.org/10.1103/PhysRevLett.83.2108} {\bibfield  {journal} {\bibinfo  {journal} {\prl}\ }\textbf {\bibinfo {volume} {83}},\ \bibinfo {pages} {2108} (\bibinfo {year} {1999})},\ \Eprint {https://arxiv.org/abs/gr-qc/9810044} {arXiv:gr-qc/9810044 [gr-qc]} \BibitemShut {NoStop}%
\bibitem [{\citenamefont {Albert}\ \emph {et~al.}(2008)\citenamefont {Albert} \emph {et~al.}}]{Albert_2008}%
  \BibitemOpen
  \bibfield  {author} {\bibinfo {author} {\bibfnamefont {J.}~\bibnamefont {Albert}} \emph {et~al.},\ }\bibfield  {title} {\bibinfo {title} {{Probing quantum gravity using photons from a flare of the active galactic nucleus Markarian 501 observed by the {MAGIC} telescope}},\ }\href {https://doi.org/10.1016/j.physletb.2008.08.053} {\bibfield  {journal} {\bibinfo  {journal} {Physics Letters B}\ }\textbf {\bibinfo {volume} {668}},\ \bibinfo {pages} {253–257} (\bibinfo {year} {2008})}\BibitemShut {NoStop}%
\bibitem [{\citenamefont {Aharonian}\ \emph {et~al.}(2008)\citenamefont {Aharonian} \emph {et~al.}}]{Aharonian_2008}%
  \BibitemOpen
  \bibfield  {author} {\bibinfo {author} {\bibfnamefont {F.}~\bibnamefont {Aharonian}} \emph {et~al.},\ }\bibfield  {title} {\bibinfo {title} {{Limits on an Energy Dependence of the Speed of Light from a Flare of the Active Galaxy {PKS} 2155-304}},\ }\bibfield  {journal} {\bibinfo  {journal} {Physical Review Letters}\ }\textbf {\bibinfo {volume} {101}},\ \href {https://doi.org/10.1103/physrevlett.101.170402} {10.1103/physrevlett.101.170402} (\bibinfo {year} {2008})\BibitemShut {NoStop}%
\bibitem [{\citenamefont {Abramowski}\ \emph {et~al.}(2011)\citenamefont {Abramowski} \emph {et~al.}}]{Abramowski_2011}%
  \BibitemOpen
  \bibfield  {author} {\bibinfo {author} {\bibfnamefont {A.}~\bibnamefont {Abramowski}} \emph {et~al.},\ }\bibfield  {title} {\bibinfo {title} {{Search for Lorentz Invariance breaking with a likelihood fit of the {PKS} 2155-304 flare data taken on {MJD} 53944}},\ }\href {https://doi.org/10.1016/j.astropartphys.2011.01.007} {\bibfield  {journal} {\bibinfo  {journal} {Astroparticle Physics}\ }\textbf {\bibinfo {volume} {34}},\ \bibinfo {pages} {738} (\bibinfo {year} {2011})}\BibitemShut {NoStop}%
\bibitem [{\citenamefont {Ellis}\ \emph {et~al.}(2003)\citenamefont {Ellis}, \citenamefont {Mavromatos}, \citenamefont {Nanopoulos},\ and\ \citenamefont {Sakharov}}]{Ellis_2003}%
  \BibitemOpen
  \bibfield  {author} {\bibinfo {author} {\bibfnamefont {J.}~\bibnamefont {Ellis}}, \bibinfo {author} {\bibfnamefont {N.~E.}\ \bibnamefont {Mavromatos}}, \bibinfo {author} {\bibfnamefont {D.~V.}\ \bibnamefont {Nanopoulos}},\ and\ \bibinfo {author} {\bibfnamefont {A.~S.}\ \bibnamefont {Sakharov}},\ }\bibfield  {title} {\bibinfo {title} {{Quantum-gravity analysis of gamma-ray bursts using wavelets}},\ }\href {https://doi.org/10.1051/0004-6361:20030263} {\bibfield  {journal} {\bibinfo  {journal} {Astronomy amp; Astrophysics}\ }\textbf {\bibinfo {volume} {402}},\ \bibinfo {pages} {409–424} (\bibinfo {year} {2003})}\BibitemShut {NoStop}%
\bibitem [{\citenamefont {Chrétien}\ \emph {et~al.}(2015)\citenamefont {Chrétien}, \citenamefont {Bolmont},\ and\ \citenamefont {Jacholkowska}}]{chrétien2015constraining}%
  \BibitemOpen
  \bibfield  {author} {\bibinfo {author} {\bibfnamefont {M.}~\bibnamefont {Chrétien}}, \bibinfo {author} {\bibfnamefont {J.}~\bibnamefont {Bolmont}},\ and\ \bibinfo {author} {\bibfnamefont {A.}~\bibnamefont {Jacholkowska}},\ }\href@noop {} {\bibinfo {title} {Constraining photon dispersion relations from observations of the vela pulsar with {H.E.S.S}}} (\bibinfo {year} {2015}),\ \Eprint {https://arxiv.org/abs/1509.03545} {arXiv:1509.03545 [astro-ph.HE]} \BibitemShut {NoStop}%
\bibitem [{\citenamefont {Ahnen}\ \emph {et~al.}(2017)\citenamefont {Ahnen} \emph {et~al.}}]{2017}%
  \BibitemOpen
  \bibfield  {author} {\bibinfo {author} {\bibfnamefont {M.~L.}\ \bibnamefont {Ahnen}} \emph {et~al.},\ }\bibfield  {title} {\bibinfo {title} {{Constraining Lorentz Invariance Violation Using the Crab Pulsar Emission Observed up to TeV Energies by {MAGIC}}},\ }\href {https://doi.org/10.3847/1538-4365/aa8404} {\bibfield  {journal} {\bibinfo  {journal} {The Astrophysical Journal Supplement Series}\ }\textbf {\bibinfo {volume} {232}},\ \bibinfo {pages} {9} (\bibinfo {year} {2017})}\BibitemShut {NoStop}%
\bibitem [{\citenamefont {Piran}\ and\ \citenamefont {Ofengeim}(2023)}]{piran2023lorentz}%
  \BibitemOpen
  \bibfield  {author} {\bibinfo {author} {\bibfnamefont {T.}~\bibnamefont {Piran}}\ and\ \bibinfo {author} {\bibfnamefont {D.~D.}\ \bibnamefont {Ofengeim}},\ }\href@noop {} {\bibinfo {title} {{Lorentz Invariance Violation Limits from {GRB} {221009A}}}} (\bibinfo {year} {2023}),\ \Eprint {https://arxiv.org/abs/2308.03031} {arXiv:2308.03031 [astro-ph.HE]} \BibitemShut {NoStop}%
\bibitem [{\citenamefont {Pasumarti}\ and\ \citenamefont {Desai}(2023)}]{Pasumarti_2023}%
  \BibitemOpen
  \bibfield  {author} {\bibinfo {author} {\bibfnamefont {V.}~\bibnamefont {Pasumarti}}\ and\ \bibinfo {author} {\bibfnamefont {S.}~\bibnamefont {Desai}},\ }\bibfield  {title} {\bibinfo {title} {{Bayesian evidence for spectral lag transition due to Lorentz invariance violation for 32 Fermi/GBM Gamma-ray bursts}},\ }\href {https://doi.org/10.1016/j.jheap.2023.10.001} {\bibfield  {journal} {\bibinfo  {journal} {Journal of High Energy Astrophysics}\ }\textbf {\bibinfo {volume} {40}},\ \bibinfo {pages} {41–48} (\bibinfo {year} {2023})}\BibitemShut {NoStop}%
\bibitem [{\citenamefont {Cao}\ \emph {et~al.}(2022)\citenamefont {Cao} \emph {et~al.}}]{LHAASO:2021opi}%
  \BibitemOpen
  \bibfield  {author} {\bibinfo {author} {\bibfnamefont {Z.}~\bibnamefont {Cao}} \emph {et~al.} (\bibinfo {collaboration} {LHAASO}),\ }\bibfield  {title} {\bibinfo {title} {{Exploring Lorentz Invariance Violation from Ultrahigh-Energy \ensuremath{\gamma} Rays Observed by LHAASO}},\ }\href {https://doi.org/10.1103/PhysRevLett.128.051102} {\bibfield  {journal} {\bibinfo  {journal} {Phys. Rev. Lett.}\ }\textbf {\bibinfo {volume} {128}},\ \bibinfo {pages} {051102} (\bibinfo {year} {2022})},\ \Eprint {https://arxiv.org/abs/2106.12350} {arXiv:2106.12350 [astro-ph.HE]} \BibitemShut {NoStop}%
\bibitem [{\citenamefont {Astapov}\ \emph {et~al.}(2019)\citenamefont {Astapov}, \citenamefont {Kirpichnikov},\ and\ \citenamefont {Satunin}}]{Astapov_2019}%
  \BibitemOpen
  \bibfield  {author} {\bibinfo {author} {\bibfnamefont {K.}~\bibnamefont {Astapov}}, \bibinfo {author} {\bibfnamefont {D.}~\bibnamefont {Kirpichnikov}},\ and\ \bibinfo {author} {\bibfnamefont {P.}~\bibnamefont {Satunin}},\ }\bibfield  {title} {\bibinfo {title} {{Photon splitting constraint on Lorentz invariance violation from Crab Nebula spectrum}},\ }\href {https://doi.org/10.1088/1475-7516/2019/04/054} {\bibfield  {journal} {\bibinfo  {journal} {Journal of Cosmology and Astroparticle Physics}\ }\textbf {\bibinfo {volume} {2019}}\bibinfo  {number} { (04)},\ \bibinfo {pages} {054–054}}\BibitemShut {NoStop}%
\bibitem [{\citenamefont {Albert}\ \emph {et~al.}(2020)\citenamefont {Albert} \emph {et~al.}}]{Albert_2020}%
  \BibitemOpen
\bibfield  {number} {  }\bibfield  {author} {\bibinfo {author} {\bibfnamefont {A.}~\bibnamefont {Albert}} \emph {et~al.},\ }\bibfield  {title} {\bibinfo {title} {{Constraints on Lorentz Invariance Violation from {HAWC} Observations of Gamma Rays above 100 {TeV}}},\ }\bibfield  {journal} {\bibinfo  {journal} {Physical Review Letters}\ }\textbf {\bibinfo {volume} {124}},\ \href {https://doi.org/10.1103/physrevlett.124.131101} {10.1103/physrevlett.124.131101} (\bibinfo {year} {2020})\BibitemShut {NoStop}%
\bibitem [{\citenamefont {Lang}\ \emph {et~al.}(2019)\citenamefont {Lang}, \citenamefont {Martínez-Huerta},\ and\ \citenamefont {de~Souza}}]{Lang_2019}%
  \BibitemOpen
  \bibfield  {author} {\bibinfo {author} {\bibfnamefont {R.~G.}\ \bibnamefont {Lang}}, \bibinfo {author} {\bibfnamefont {H.}~\bibnamefont {Martínez-Huerta}},\ and\ \bibinfo {author} {\bibfnamefont {V.}~\bibnamefont {de~Souza}},\ }\bibfield  {title} {\bibinfo {title} {{Improved limits on Lorentz invariance violation from astrophysical gamma-ray sources}},\ }\bibfield  {journal} {\bibinfo  {journal} {Physical Review D}\ }\textbf {\bibinfo {volume} {99}},\ \href {https://doi.org/10.1103/physrevd.99.043015} {10.1103/physrevd.99.043015} (\bibinfo {year} {2019})\BibitemShut {NoStop}%
\bibitem [{\citenamefont {Abdalla}\ \emph {et~al.}(2019)\citenamefont {Abdalla} \emph {et~al.}}]{Abdalla_2019}%
  \BibitemOpen
  \bibfield  {author} {\bibinfo {author} {\bibfnamefont {H.}~\bibnamefont {Abdalla}} \emph {et~al.},\ }\bibfield  {title} {\bibinfo {title} {{The 2014 TeV $\gamma$-Ray Flare of {Mrk} 501 Seen with {H.E.S.S.}: Temporal and Spectral Constraints on Lorentz Invariance Violation}},\ }\href {https://doi.org/10.3847/1538-4357/aaf1c4} {\bibfield  {journal} {\bibinfo  {journal} {The Astrophysical Journal}\ }\textbf {\bibinfo {volume} {870}},\ \bibinfo {pages} {93} (\bibinfo {year} {2019})}\BibitemShut {NoStop}%
\bibitem [{\citenamefont {Li}\ and\ \citenamefont {Ma}(2023)}]{Li_2023}%
  \BibitemOpen
  \bibfield  {author} {\bibinfo {author} {\bibfnamefont {H.}~\bibnamefont {Li}}\ and\ \bibinfo {author} {\bibfnamefont {B.-Q.}\ \bibnamefont {Ma}},\ }\bibfield  {title} {\bibinfo {title} {Revisiting lorentz invariance violation from {GRB} {221009A}},\ }\href {https://doi.org/10.1088/1475-7516/2023/10/061} {\bibfield  {journal} {\bibinfo  {journal} {Journal of Cosmology and Astroparticle Physics}\ }\textbf {\bibinfo {volume} {2023}}\bibinfo  {number} { (10)},\ \bibinfo {pages} {061}}\BibitemShut {NoStop}%
\bibitem [{\citenamefont {de~los Heros}\ and\ \citenamefont {Terzić}(2022)}]{heros2022cosmic}%
  \BibitemOpen
\bibfield  {number} {  }\bibfield  {author} {\bibinfo {author} {\bibfnamefont {C.~P.}\ \bibnamefont {de~los Heros}}\ and\ \bibinfo {author} {\bibfnamefont {T.}~\bibnamefont {Terzić}},\ }\href@noop {} {\bibinfo {title} {{Cosmic searches for Lorentz invariance violation}}} (\bibinfo {year} {2022}),\ \Eprint {https://arxiv.org/abs/2209.06531} {arXiv:2209.06531 [astro-ph.HE]} \BibitemShut {NoStop}%
\bibitem [{\citenamefont {Amelino-Camelia}\ \emph {et~al.}(1998)\citenamefont {Amelino-Camelia}, \citenamefont {Ellis}, \citenamefont {Mavromatos}, \citenamefont {Nanopoulos},\ and\ \citenamefont {Sarkar}}]{Amelino_Camelia_1998}%
  \BibitemOpen
  \bibfield  {author} {\bibinfo {author} {\bibfnamefont {G.}~\bibnamefont {Amelino-Camelia}}, \bibinfo {author} {\bibfnamefont {J.}~\bibnamefont {Ellis}}, \bibinfo {author} {\bibfnamefont {N.~E.}\ \bibnamefont {Mavromatos}}, \bibinfo {author} {\bibfnamefont {D.~V.}\ \bibnamefont {Nanopoulos}},\ and\ \bibinfo {author} {\bibfnamefont {S.}~\bibnamefont {Sarkar}},\ }\bibfield  {title} {\bibinfo {title} {{Tests of quantum gravity from observations of $\gamma$-ray bursts}},\ }\href {https://doi.org/10.1038/31647} {\bibfield  {journal} {\bibinfo  {journal} {Nature}\ }\textbf {\bibinfo {volume} {393}},\ \bibinfo {pages} {763} (\bibinfo {year} {1998})}\BibitemShut {NoStop}%
\bibitem [{\citenamefont {Vasileiou}\ \emph {et~al.}(2013)\citenamefont {Vasileiou} \emph {et~al.}}]{Vasileiou_2013}%
  \BibitemOpen
  \bibfield  {author} {\bibinfo {author} {\bibfnamefont {V.}~\bibnamefont {Vasileiou}} \emph {et~al.},\ }\bibfield  {title} {\bibinfo {title} {{Constraints on Lorentz invariance violation from Fermi-Large Area Telescope observations of gamma-ray bursts}},\ }\bibfield  {journal} {\bibinfo  {journal} {Physical Review D}\ }\textbf {\bibinfo {volume} {87}},\ \href {https://doi.org/10.1103/physrevd.87.122001} {10.1103/physrevd.87.122001} (\bibinfo {year} {2013})\BibitemShut {NoStop}%
\bibitem [{\citenamefont {Acciari}\ and\ \citenamefont {other}(2020)}]{Acciari_2020}%
  \BibitemOpen
  \bibfield  {author} {\bibinfo {author} {\bibfnamefont {V.}~\bibnamefont {Acciari}}\ and\ \bibinfo {author} {\bibnamefont {other}},\ }\bibfield  {title} {\bibinfo {title} {{Bounds on Lorentz Invariance Violation from {MAGIC} Observation of {GRB} {190114C}}},\ }\bibfield  {journal} {\bibinfo  {journal} {Physical Review Letters}\ }\textbf {\bibinfo {volume} {125}},\ \href {https://doi.org/10.1103/physrevlett.125.021301} {10.1103/physrevlett.125.021301} (\bibinfo {year} {2020})\BibitemShut {NoStop}%
\bibitem [{\citenamefont {Li}\ \emph {et~al.}(2004)\citenamefont {Li}, \citenamefont {Qu}, \citenamefont {Feng}, \citenamefont {Song}, \citenamefont {Ding},\ and\ \citenamefont {Chen}}]{Li_2004}%
  \BibitemOpen
  \bibfield  {author} {\bibinfo {author} {\bibfnamefont {T.-P.}\ \bibnamefont {Li}}, \bibinfo {author} {\bibfnamefont {J.-L.}\ \bibnamefont {Qu}}, \bibinfo {author} {\bibfnamefont {H.}~\bibnamefont {Feng}}, \bibinfo {author} {\bibfnamefont {L.-M.}\ \bibnamefont {Song}}, \bibinfo {author} {\bibfnamefont {G.-Q.}\ \bibnamefont {Ding}},\ and\ \bibinfo {author} {\bibfnamefont {L.}~\bibnamefont {Chen}},\ }\bibfield  {title} {\bibinfo {title} {{Timescale Analysis of Spectral Lags}},\ }\href {https://doi.org/10.1088/1009-9271/4/6/583} {\bibfield  {journal} {\bibinfo  {journal} {Chinese Journal of Astronomy and Astrophysics}\ }\textbf {\bibinfo {volume} {4}},\ \bibinfo {pages} {583–598} (\bibinfo {year} {2004})}\BibitemShut {NoStop}%
\bibitem [{\citenamefont {Mart{\'{\i}}nez}\ and\ \citenamefont {Errando}(2009)}]{Mart_nez_2009}%
  \BibitemOpen
  \bibfield  {author} {\bibinfo {author} {\bibfnamefont {M.}~\bibnamefont {Mart{\'{\i}}nez}}\ and\ \bibinfo {author} {\bibfnamefont {M.}~\bibnamefont {Errando}},\ }\bibfield  {title} {\bibinfo {title} {A new approach to study energy-dependent arrival delays on photons from astrophysical sources},\ }\href {https://doi.org/10.1016/j.astropartphys.2009.01.005} {\bibfield  {journal} {\bibinfo  {journal} {Astroparticle Physics}\ }\textbf {\bibinfo {volume} {31}},\ \bibinfo {pages} {226} (\bibinfo {year} {2009})}\BibitemShut {NoStop}%
\bibitem [{\citenamefont {Cao}\ \emph {et~al.}(2023{\natexlab{a}})\citenamefont {Cao} \emph {et~al.}}]{2023}%
  \BibitemOpen
  \bibfield  {author} {\bibinfo {author} {\bibfnamefont {Z.}~\bibnamefont {Cao}} \emph {et~al.},\ }\bibfield  {title} {\bibinfo {title} {A tera{\textendash}electron volt afterglow from a narrow jet in an extremely bright gamma-ray burst},\ }\href {https://doi.org/10.1126/science.adg9328} {\bibfield  {journal} {\bibinfo  {journal} {Science}\ }\textbf {\bibinfo {volume} {380}},\ \bibinfo {pages} {1390} (\bibinfo {year} {2023}{\natexlab{a}})}\BibitemShut {NoStop}%
\bibitem [{\citenamefont {Cao}\ \emph {et~al.}(2023{\natexlab{b}})\citenamefont {Cao} \emph {et~al.}}]{cao2023high}%
  \BibitemOpen
  \bibfield  {author} {\bibinfo {author} {\bibfnamefont {Z.}~\bibnamefont {Cao}} \emph {et~al.},\ }\bibfield  {title} {\bibinfo {title} {{Very high-energy gamma-ray emission beyond 10 {TeV} from {GRB}{221009A}}},\ }\href {https://doi.org/10.1126/sciadv.adj2778} {\bibfield  {journal} {\bibinfo  {journal} {Science Advances}\ }\textbf {\bibinfo {volume} {9}},\ \bibinfo {pages} {eadj2778} (\bibinfo {year} {2023}{\natexlab{b}})},\ \Eprint {https://arxiv.org/abs/https://www.science.org/doi/pdf/10.1126/sciadv.adj2778} {https://www.science.org/doi/pdf/10.1126/sciadv.adj2778} \BibitemShut {NoStop}%
\bibitem [{\citenamefont {Aghanim}\ \emph {et~al.}(2020)\citenamefont {Aghanim} \emph {et~al.}}]{Planck_2018}%
  \BibitemOpen
  \bibfield  {author} {\bibinfo {author} {\bibfnamefont {N.}~\bibnamefont {Aghanim}} \emph {et~al.},\ }\bibfield  {title} {\bibinfo {title} {{Planck 2018 results: {VI}. Cosmological parameters}},\ }\href {https://doi.org/10.1051/0004-6361/201833910} {\bibfield  {journal} {\bibinfo  {journal} {Astronomy amp; Astrophysics}\ }\textbf {\bibinfo {volume} {641}},\ \bibinfo {pages} {A6} (\bibinfo {year} {2020})}\BibitemShut {NoStop}%
\bibitem [{\citenamefont {Terzi{\'{c}}}\ \emph {et~al.}(2021)\citenamefont {Terzi{\'{c}}}, \citenamefont {Kerszberg},\ and\ \citenamefont {Stri{\v{s}}kovi{\'{c}}}}]{Terzi__2021}%
  \BibitemOpen
  \bibfield  {author} {\bibinfo {author} {\bibfnamefont {T.}~\bibnamefont {Terzi{\'{c}}}}, \bibinfo {author} {\bibfnamefont {D.}~\bibnamefont {Kerszberg}},\ and\ \bibinfo {author} {\bibfnamefont {J.}~\bibnamefont {Stri{\v{s}}kovi{\'{c}}}},\ }\bibfield  {title} {\bibinfo {title} {{Probing Quantum Gravity with Imaging Atmospheric Cherenkov Telescopes}},\ }\href {https://doi.org/10.3390/universe7090345} {\bibfield  {journal} {\bibinfo  {journal} {Universe}\ }\textbf {\bibinfo {volume} {7}},\ \bibinfo {pages} {345} (\bibinfo {year} {2021})}\BibitemShut {NoStop}%
\bibitem [{\citenamefont {{Veres}}\ \emph {et~al.}(2022)\citenamefont {{Veres}}, \citenamefont {{Burns}}, \citenamefont {{Bissaldi}}, \citenamefont {{Lesage}}, \citenamefont {{Roberts}},\ and\ \citenamefont {{Fermi GBM Team}}}]{2022GCN.32636....1V}%
  \BibitemOpen
  \bibfield  {author} {\bibinfo {author} {\bibfnamefont {P.}~\bibnamefont {{Veres}}}, \bibinfo {author} {\bibfnamefont {E.}~\bibnamefont {{Burns}}}, \bibinfo {author} {\bibfnamefont {E.}~\bibnamefont {{Bissaldi}}}, \bibinfo {author} {\bibfnamefont {S.}~\bibnamefont {{Lesage}}}, \bibinfo {author} {\bibfnamefont {O.}~\bibnamefont {{Roberts}}},\ and\ \bibinfo {author} {\bibnamefont {{Fermi GBM Team}}},\ }\bibfield  {title} {\bibinfo {title} {{GRB 221009A: Fermi GBM detection of an extraordinarily bright GRB}},\ }\href@noop {} {\bibfield  {journal} {\bibinfo  {journal} {GRB Coordinates Network}\ }\textbf {\bibinfo {volume} {32636}},\ \bibinfo {pages} {1} (\bibinfo {year} {2022})}\BibitemShut {NoStop}%
\bibitem [{\citenamefont {{Lesage}}\ \emph {et~al.}(2023)\citenamefont {{Lesage}} \emph {et~al.}}]{2023ApJ...952L..42L}%
  \BibitemOpen
  \bibfield  {author} {\bibinfo {author} {\bibfnamefont {S.}~\bibnamefont {{Lesage}}} \emph {et~al.},\ }\bibfield  {title} {\bibinfo {title} {{Fermi-GBM Discovery of GRB 221009A: An Extraordinarily Bright GRB from Onset to Afterglow}},\ }\bibfield  {journal} {\bibinfo  {journal} {\apj}\ }\textbf {\bibinfo {volume} {952}},\ \href {https://doi.org/10.3847/2041-8213/ace5b4} {10.3847/2041-8213/ace5b4} (\bibinfo {year} {2023}),\ \Eprint {https://arxiv.org/abs/2303.14172} {2303.14172} \BibitemShut {NoStop}%
\bibitem [{\citenamefont {{de Ugarte Postigo}}\ \emph {et~al.}(2022)\citenamefont {{de Ugarte Postigo}} \emph {et~al.}}]{2022GCN.32648....1D}%
  \BibitemOpen
  \bibfield  {author} {\bibinfo {author} {\bibfnamefont {A.}~\bibnamefont {{de Ugarte Postigo}}} \emph {et~al.},\ }\bibfield  {title} {\bibinfo {title} {{GRB 221009A: Redshift from X-shooter/VLT}},\ }\href@noop {} {\bibfield  {journal} {\bibinfo  {journal} {GRB Coordinates Network}\ }\textbf {\bibinfo {volume} {32648}},\ \bibinfo {pages} {1} (\bibinfo {year} {2022})}\BibitemShut {NoStop}%
\bibitem [{\citenamefont {{Castro-Tirado}}\ \emph {et~al.}(2022)\citenamefont {{Castro-Tirado}} \emph {et~al.}}]{2022GCN.32686....1C}%
  \BibitemOpen
  \bibfield  {author} {\bibinfo {author} {\bibfnamefont {A.~J.}\ \bibnamefont {{Castro-Tirado}}} \emph {et~al.},\ }\bibfield  {title} {\bibinfo {title} {{GRB 221009A: 10.4m GTC spectroscopic redshift confirmation}},\ }\href@noop {} {\bibfield  {journal} {\bibinfo  {journal} {GRB Coordinates Network}\ }\textbf {\bibinfo {volume} {32686}},\ \bibinfo {pages} {1} (\bibinfo {year} {2022})}\BibitemShut {NoStop}%
\bibitem [{\citenamefont {Abramowski}\ \emph {et~al.}(2015)\citenamefont {Abramowski} \emph {et~al.}}]{Abramowski_2015}%
  \BibitemOpen
  \bibfield  {author} {\bibinfo {author} {\bibfnamefont {A.}~\bibnamefont {Abramowski}} \emph {et~al.},\ }\bibfield  {title} {\bibinfo {title} {{The} 2012 {flare} {of} {PG} 1553-113 {seen} {with} {H.E.S.S.} {and} {Fermi}-{LAT}:{Constraints} on the source redshift and {Lorentz} invariance violation},\ }\href {https://doi.org/10.1088/0004-637x/802/1/65} {\bibfield  {journal} {\bibinfo  {journal} {The Astrophysical Journal}\ }\textbf {\bibinfo {volume} {802}},\ \bibinfo {pages} {65} (\bibinfo {year} {2015})}\BibitemShut {NoStop}%
\bibitem [{\citenamefont {Saldana-Lopez}\ \emph {et~al.}(2021)\citenamefont {Saldana-Lopez} \emph {et~al.}}]{Saldana_Lopez_2021}%
  \BibitemOpen
  \bibfield  {author} {\bibinfo {author} {\bibfnamefont {A.}~\bibnamefont {Saldana-Lopez}} \emph {et~al.},\ }\bibfield  {title} {\bibinfo {title} {{An observational determination of the evolving extragalactic background light from the multiwavelength {HST}/{CANDELS} survey in the Fermi and {CTA} era}},\ }\href {https://doi.org/10.1093/mnras/stab2393} {\bibfield  {journal} {\bibinfo  {journal} {Monthly Notices of the Royal Astronomical Society}\ }\textbf {\bibinfo {volume} {507}},\ \bibinfo {pages} {5144} (\bibinfo {year} {2021})}\BibitemShut {NoStop}%
\end{thebibliography}%
\begin{appendix}

\section{Systematic Errors\label{SE}}
The fitting parameters of the light curve are subject to uncertainties, especially during the rapid rise phase, where the fitting uncertainty is relatively high due to the low number of observed photons. In this appendix, we investigate the impact of these fitting uncertainties on the constraint results. As previously discussed, the constraints on the LIV energy scale $E_{\text{QG},1}$($E_{\text{QG},2}$) is roughly proportional to $t^{-1}_\text{var}$ ($t^{-1/2}_\text{var}$). We examine the two most extreme cases by selecting the parameters $\left\{\omega_1,\alpha_0,\alpha_1,\alpha_2,\alpha_3,t_{b,1},t_{b,2}\right\}$ ($t_{b,0}$ fixed) within the $1\sigma$ deviation ranges. These two cases with the parameter sets $I_f = \{2.11, 22.3, 2.16, -1.139, -2.19, 14.3, 460\}$ and $I_s =\{1, 7.4, 1.34, -1.079, -1.64, 17.6, 720\}$ correspond to the fastest and slowest variability, respectively. The constraints on the QG energy scale from Model-1 with the two parameter sets are listed in Tab.~\ref{Table4}. For the purpose of facilitating comparison, the constraints obtained with the best-fit parameters $I_b$ are also listed in this table.
\begin{table}[htbp]
    \centering
    \caption{The 95$\%$ CL lower limits on the QG energy from Model-1 with the values of parameters chosen as $I_f$, $I_b$ and $I_s$.}
    \begin{ruledtabular}
    \begin{tabular}{c|cc|cc}
    &\multicolumn{2}{c|}{$E_{\text{QG},1}\left[10^{19}\text{GeV}\right]$}&\multicolumn{2}{c}{$E_{\text{QG},2}\left[10^{11}\text{GeV}\right]$}\\
    &$\mathcal{S}=+1$&$\mathcal{S}=-1$& $\mathcal{S}=+1$&$\mathcal{S}=-1$\\
    \hline
    $I_f$&16.5&11.6&12.5&10.6\\
    $I_b$&14.7&6.5&12.0&7.2\\
    $I_s$&11.0&3.9&11.0&5.6\\
    \end{tabular}
    \end{ruledtabular}
    \label{Table4}
    \end{table}

As indicated in Tab.~\ref{Table4}, the constraints on the subluminal (superluminal) scenario are enhanced with $I_f$ and weakened with $I_s$ compared to the result with $I_b$ for both $n=1$ and $n=2$. Concretely, for $n=1$ the constraints are enhanced and weakened by a factor of $1.12$ ($1.78$) and $1.34$ ($1.67$) for the subluminal (superluminal) scenario. According to the coarse relations  $E_{\text{QG},1}\propto t^{-1}_\text{var}$ and $E_{\text{QG},2}\propto t^{-1/2}_\text{var}$, the constraints in the $n=2$ case would be enhanced and weakened by a factor of $\sqrt{1.12}$ ($\sqrt{1.78}$) and $\sqrt{1.34}$ ($\sqrt{1.67}$) for the subluminal (superluminal) scenario, respectively. These rough estimates are consistent with our actual results.

\section{Calibrated Confidence Intervals\label{CI}}
We follow the method outlined in \cite{Acciari_2020} to construct the 95$\%$ calibrated CIs using Monte Carlo simulations. To accomplish this, we generate 1000 sets of mock data, each comprising 142 photons, with statistical properties similar to the actual data. The construction of each data set proceeds as follows. Firstly, 61 (64) random arrival times within 230-300 s (300-900 s) are sampled according to the form of the light curve $\lambda (t)$. Then, 61 (64) energies in the range of 3-13 TeV are sampled based on the convolution of the KM2A's effective area and the power-law spectral form with an exponential cutoff, which can fit the observed spectrum of LHAASO \cite{cao2023high}. Subsequently, the sampled times and energies are combined randomly to obtain the arrival times and estimated energies of 125 signal photons. Additionally, 3 and 14 background events are sampled according to the background PDF $f_b(E_\text{est})$ within the two time intervals, respectively.

\begin{figure}[htbp]
    \centering
    \includegraphics[width=\linewidth]{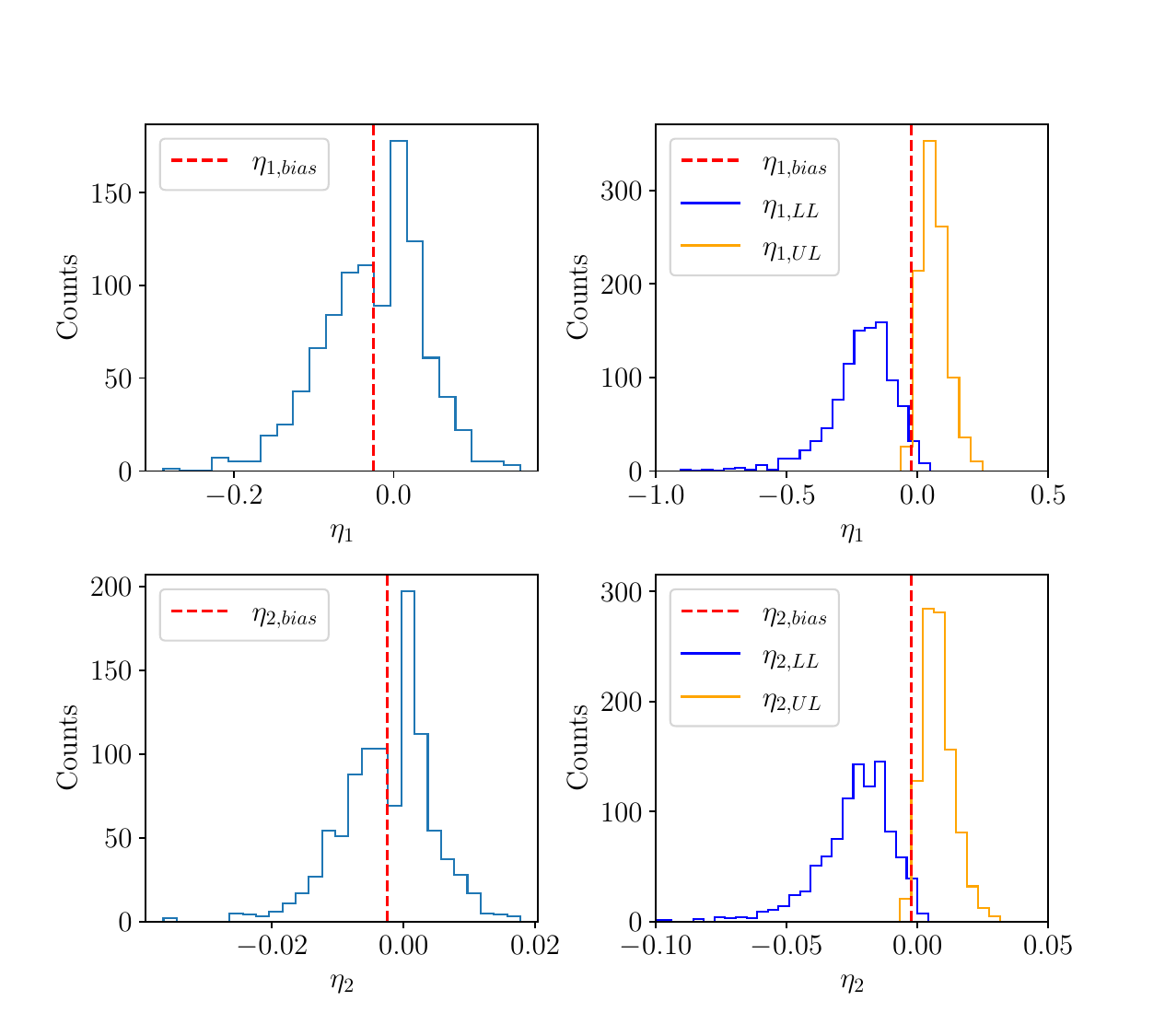}
    \caption{Left: The distributions of $\hat{\eta}^{(i)}$ from the mock data; Right: The distributions of $\eta^{(i)}_{\text{LL}}$ and $\eta^{(i)}_{\text{UL}}$ from the mock data for $L_l$ and $L_r$, respectively. The red dashed lines represent the positions of $\eta_\text{bias}$. Top and bottom panels represent the n=1 and n=2 cases, respectively. All these results are obtained with Model-1.}
    \label{MC1}
\end{figure}

The ML analysis is then performed on these 1000 mock data sets, resulting in 1000 $L(\eta_1)$ and 1000 $L(\eta_2)$ curves. The discussion below applies to both n = 1 and 2, so the subscripts 1 and 2 are omitted. The 1000 curves have 1000 minima, denoted as $\{\hat{\eta}^{(i)} \left|i=1,\cdots,1000\right.\}$. As there is no correlation between the arrival times and estimated energies of the randomly sampled data, it can be expected that no LIV effect is present among them in the mock data. Therefore, the minimum of $L$ should distribute around zero, with a mean value equal to zero. However, similar to the PV analysis, the mean value of $\hat{\eta}^{(i)}$ slightly deviates from zero, which is attributed to the inherent bias of the ML method, denoted as $\eta_\text{bias}$. The calibrated best value is obtained by subtracting the above bias from the best fit value $\hat{\eta}$, which is derived from the ML analysis for the real data, denoted as
$\eta^{\text{cal}}_{\text{best}}=\hat{\eta}-\eta_{\text{bias}}
$.

\begin{table}[htbp]
    \centering
    \caption{$\eta_\text{bias}$, $L_l$, and $L_r$ of the three models obtained from Monte Carlo simulations.}
    \begin{ruledtabular}
    \begin{tabular}{c|cc|cc|cc}
    &\multicolumn{2}{c|}{Model-1}&\multicolumn{2}{c|}{Model-2}&\multicolumn{2}{c}{Model-3}\\
    &$n=1$& $n=2$&$n=1$&$n=2$& $n=1$&$n=2$\\
    \hline
    $\eta_\text{bias}$&-0.025 &-0.003&0.006&0.001&-0.001&-0.001\\
    $L_l$&3.55&3.51 &3.21&3.28&4.16&4.52\\
    $L_r$&3.33 &3.21 &5.21&4.80&3.78&2.91\\
    \end{tabular}
    \end{ruledtabular}
    \label{Table3}
    \end{table}
    
The lower and upper bounds of the 95$\%$ CI of $\eta$, denoted as $\eta_\text{LL}$ and $\eta_\text{UL}$, correspond to the left and right thresholds $L_l$ and $L_r$ on the $L(\eta)$ curve, respectively. Here $L(\eta)$ is obtained from the real data. The calibrated bounds are obtained by subtracting the bias as follows:
\begin{equation}
    \begin{aligned}
        &\eta^{\text{cal}}_{\text{LL}}=\eta_{\text{LL}}-\eta_{\text{bias}},\\
        &\eta^{\text{cal}}_{\text{UL}}=\eta_{\text{UL}}-\eta_{\text{bias}}.
    \end{aligned}
\end{equation}

The thresholds $L_l$ and $L_r$ are obtained as follows. For a given threshold value $\tilde{L}_l$, we can obtain a $\tilde{\eta}^{(i)}_{\text{LL}}$ 
by requiring $L^{(i)}(\tilde{\eta}^{(i)}_{\text{LL}})= \tilde{L}_l$ for the $i$-th mock data set, where $L^{(i)}(\eta)$ is the $L$ function of this data set. 
Subsequently, we obtain the distribution of ${\tilde{\eta}^{(i)}_{\text{LL}}}$ from all the mock data sets for the given $\tilde{L}_l$, and count the number $N(\tilde{L}_l)$ of $\tilde{\eta}_\text{LL}$ that are less than $\eta_\text{bias}$ according to this distribution. Finally, $L_l$ is derived by finding the distribution satisfying with $N(L_l)=975$. This means that the probability of finding a $\tilde{\eta}_\text{LL}$ less than $\eta_\text{bias}$ is $97.5\%$ in this distribution. Using the similar method, we can also obtain $L_r$.

The distribution of the $\hat{\eta}^{(i)}$, $\eta^{(i)}_\text{LL}$ corresponding to $L_l$, and $\eta^{(i)}_\text{UL}$ corresponding to $L_r$ for Model-1 are shown in Fig.~\ref{MC1}. The values of $\eta_\text{bias},L_l$, and $L_r$ for all the three models are listed in the Tab.~\ref{Table3}.

\end{appendix}
\end{document}